\pdfoutput=1
\documentclass[aps,pra,twocolumn,superscriptaddress,showpacs,floatfix,nofootinbib,llncs,longbibliography,pdftex]{revtex4-2}
\usepackage{graphicx}
\usepackage{amssymb}
\usepackage{amsmath}
\usepackage{color}
\usepackage{psfrag}
\usepackage{epsfig}
\usepackage{bbm}
\usepackage{bm}
\usepackage{hyperref}
\usepackage[normalem]{ulem}
\usepackage{amsthm}
\usepackage{epstopdf}
\usepackage{verbatim}
\usepackage{subfigure}
\usepackage[export]{adjustbox}
\usepackage{bbold}

\usepackage{array}
\usepackage{mathtools}

\DeclareMathOperator{\tr}{tr}

\newcommand{\tot}{\text{tot}}
\newcommand{\s}{\text{S}}

\newcommand{\e}{\text{E}}
\newcommand{\PM}{\text{PM}}
\newcommand{\ADO}{\text{ADO}}
\newcommand{\ep}{\text{EP}}
\newcommand{\bfj}{\textbf{j}}
\newcommand{\bfk}{\textbf{k}}

\newcolumntype{M}[1]{>{\centering\arraybackslash}m{#1}}

\begin{document}
\title{Non-Markovian Quantum Exceptional Points
}
\author{Jhen-Dong Lin}
\email{jhendonglin@gmail.com}
\affiliation{Department of Physics, National Cheng Kung University, 701 Tainan, Taiwan}
\affiliation{Center for Quantum Frontiers of Research \& Technology, NCKU, 70101 Tainan, Taiwan}

\author{Po-Chen Kuo}
\affiliation{Department of Physics, National Cheng Kung University, 701 Tainan, Taiwan}
\affiliation{Center for Quantum Frontiers of Research \& Technology, NCKU, 70101 Tainan, Taiwan}

\author{Neill Lambert}
\affiliation{Theoretical Quantum Physics Laboratory, RIKEN Cluster for Pioneering Research, Wako-shi, Saitama 351-0198, Japan}
\affiliation{RIKEN Center for Quantum Computing (RQC), Wakoshi, Saitama 351-0198, Japan}
\author{Adam Miranowicz}
\affiliation{Institute of Spintronics and Quantum Information, Faculty of Physics, Adam Mickiewicz University, 61-614 Poznań, Poland}

\author{Franco Nori}
\email{fnori@riken.jp}
\affiliation{Theoretical Quantum Physics Laboratory, RIKEN Cluster for Pioneering Research, Wako-shi, Saitama 351-0198, Japan}
\affiliation{RIKEN Center for Quantum Computing (RQC), Wakoshi, Saitama 351-0198, Japan}
\affiliation{Quantum Research Institute, The University of Michigan, Ann Arbor, 48109-1040 Michigan, USA}
\author{Yueh-Nan Chen}
\email{yuehnan@mail.ncku.edu.tw}
\affiliation{Department of Physics, National Cheng Kung University, 701 Tainan, Taiwan}
\affiliation{Center for Quantum Frontiers of Research \& Technology, NCKU, 70101 Tainan, Taiwan}
\affiliation{Physics Division, National Center for Theoretical Sciences, Taipei 106319, Taiwan}

\date{\today}

\begin{abstract}
Exceptional points (EPs) are singularities in the spectra of non-Hermitian operators, where eigenvalues and eigenvectors coalesce. Recently, open quantum systems have been increasingly explored as EP testbeds due to their natural non-Hermitian nature. However, existing works mostly focus on the Markovian limit, leaving a gap in understanding EPs in the non-Markovian regime.
In this work, we address this gap by proposing a theoretical framework based on two numerically exact descriptions of non-Markovian dynamics: the pseudomode mapping and the hierarchical equations of motion. The proposed framework enables conventional spectral analysis for EP identification, establishing direct links between EPs and dynamic manifestations in open systems, such as non-exponential decays and enhanced sensitivity to external perturbations.
We unveil pure non-Markovian EPs that are unobservable in the Markovian limit. Remarkably, the EP aligns with the Markovian-to-non-Markovian transition, and the EP condition is adjustable by modifying environmental spectral properties. Moreover, we show that structured environments can elevate EP order, thereby enhancing the system's sensitivity. These findings lay a theoretical foundation and open new avenues for non-Markovian reservoir engineering and non-Hermitian physics.
\end{abstract}

\maketitle
\section{Introduction}
Spectral singularities for non-Hermitian systems, known as exceptional points (EPs), have attracted intense research attention over the past decades~\cite{heiss2012physics, el2018non,
 miri2019exceptional, ozdemir2019parity}. These singularities are pivotal in studying open systems, as environmental noise inherently breaks the Hermiticity. Early studies of EPs mainly focus on non-Hermitian Hamiltonians (NHHs)~\cite{PhysRevLett.80.5243, bender2007making,
moiseyev2011non, mostafazadeh2002pseudo} that are suitable for modeling classical and semiclassical systems. In these settings, (semi-)classical EPs, often termed Hamiltonian EPs (HEPs), are identified by the convergence of at least two eigenvalues and their corresponding eigenstates within NHHs. Notably, it has been shown that environmental noises can induce exotic effects near HEPs, e.g., EP-induced lasing~\cite{peng2014loss,
PhysRevLett.113.053604,
zhang2018phonon}, programmable mode switching~\cite{arkhipov2023dynamically}, and EP-enhanced sensitivity~\cite{chen2017exceptional, hodaei2017enhanced, rechtsman2017optical, wiersig2020review, PhysRevA.101.013814}. 

Recently, investigations of EPs have extended into the full quantum regime~\cite{PhysRevA.100.062131, PhysRevA.101.013812,
PhysRevA.101.062112, PhysRevA.102.033715, PhysRevA.104.012205, PRXQuantum.2.040346, zhang2022dynamical, perina2022quantum, PhysRevLett.131.260201, PhysRevResearch.5.043036, PhysRevLett.130.110402, khandelwal2023chiral, abo2024liouvillian, PhysRevA.109.033511}, where the temporal evolution of an open quantum system is governed by a Lindblad master equation or, equivalently, by a Liouvillian superoperator. Unlike the effective NHHs, Liouvillian superoperators incorporate the concept of quantum jumps into the dynamical process~\cite{breuer2002theory}. In this context, the EPs associated with Liouvillian superoperators are termed as quantum EPs or Liouvillian EPs (LEPs). It has been demonstrated that \emph{pure quantum EPs} exist~\cite{PhysRevA.100.062131}, which are phenomena without (semi-)classical counterparts.

To date, the exploration for both HEPs and LEPs has largely been confined to the memoryless Markovian limit, which is only valid in cases of sufficiently weak system-environment interaction or environments without any structure. Recently, several works have indicated that EP-like critical behaviors~\cite{PhysRevA.97.052116, PhysRevResearch.3.033029, sergeev2023signature, PhysRevA.106.053709} could manifest in the non-Markovian regime. However, whether the concepts of EPs can be directly applied to the non-Markovian regime remains an open question.
A primary challenge lies within the structure of the non-Markovian equation of motion for reduced dynamics, which can be generally expressed as:  
\begin{equation}
    \frac{d\rho_{\s}(t)}{d t} = \int_0^t d\tau~ \mathcal{K}(t,\tau) \rho_\s(\tau). \label{eq: nonmarkov dynamics with memory kernel}
\end{equation}
Here, the non-Markovian effect is encoded in the memory kernel $\mathcal{K}(t,\tau)$~\cite{breuer2002theory}, and $\rho_\s(t)$ denotes the open system's reduced density operator. The integral-differential nature of this time-non-local equation complicates the application of traditional spectral analysis techniques for identifying EPs.

In this work, we aim to address this theoretical gap. The main result lies in the development of a systematic framework for quantum EPs associated with generic non-Markovian open systems. The idea is based on applying the pseudomode equation of motion (PMEOM)~\cite{PhysRevA.55.2290, PhysRevA.64.053813, PhysRevA.80.012104, pleasance2021pseudomode, PhysRevLett.120.030402, Lambert2019, PhysRevResearch.2.043058, Mauro2023, PRXQuantum.4.030316, cirio2023modeling, menczel2024non} and the hierarchical equations of motion (HEOM)~\cite{jin2008exact, kreisbeck2011high, PhysRevA.85.062323, Yan2012, dunn2019removing, Tanimura01, ikeda2020generalization, Mauro2022, lambert2020bofinheom, Huang2023}, which can be used to describe a large class of system-environment models. In contrast to the memory kernel approach as described in Eq.~\eqref{eq: nonmarkov dynamics with memory kernel}, these methods encode the non-Markovian effects in auxiliary degrees of freedom so that both PMEOM and HEOM can be formulated by time-local differential equations. Specifically, the dynamics is governed by what we call \emph{extended Liouvillian superoperators}, enabling us to perform conventional spectral analysis, identifying the corresponding EPs, and revealing their impacts on the non-Markovian open systems.

Moreover, we demonstrate the utility through two analytically tractable examples. The first one is the spin-boson model with a Lorentzian environment. We identify a \emph{pure non-Markovian EP} that cannot be observed in the Markovian limit. Intriguingly, the EP precisely aligns with the Markovian-to-non-Markovian transition, thereby revealing a close relationship between the onset of non-Markovian information backflow and non-Hermitian phase transition. In addition, we show that the EP criterion is tunable by reservoir engineering. Specifically, we introduce a band gap to the Lorentzian environment, demonstrating that the corresponding EP condition requires a smaller system-environment coupling strength compared to the gap-less scenario.

For the second example, we examine linear bosonic systems by using the adjoint PMEOM within the Heisenberg picture. We show that the dynamics of the modes' amplitudes can be determined by effective NHHs, enabling us to explore potential non-Markovian HEPs. As an example, we consider a two-coupled-modes system, where a second-order HEP emerges in the Markovian limit. We showcase that in the non-Markovian regime, the HEP can be transformed to the third order. Consequently, the system becomes more sensitive to external perturbations with the help of the quantum memory effect. These findings reveal the intricate interplay between EPs and the memory effect, laying a theoretical foundation for exploring non-Hermitian physics towards non-Markovian and non-perturbative regimes. 

\section{Results}
\subsection*{General framework for non-Markovian exceptional points}

\subsubsection{Pseudomode equation of motion}
\begin{figure*}
\includegraphics[width = 1\linewidth]{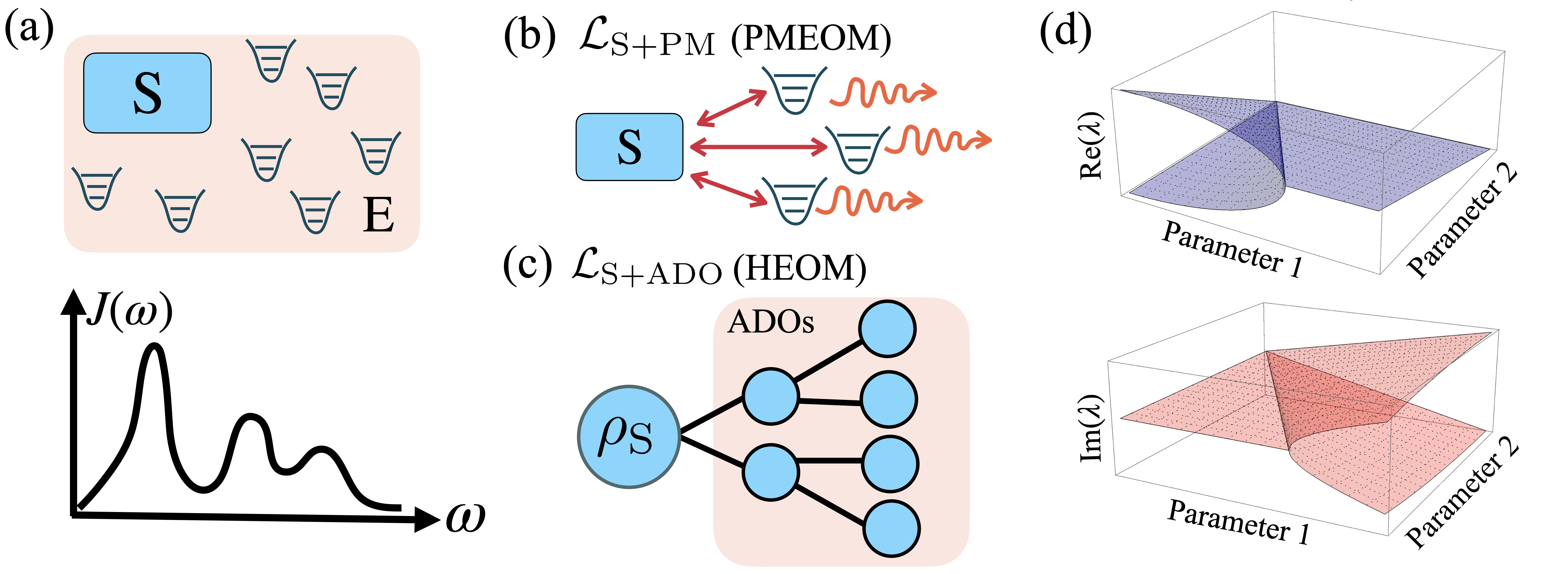}
\caption{Schematic illustration depicting EPs for (a) a generic non-Markovian open-system model, where the structured environment is captured by the spectral density function $J(\omega)$. For a given spectral density function and the corresponding environmental correlation function, the exact non-Markovian dynamics can either be described by (b) PMEOM or (c) HEOM with the corresponding extended Liouvillian superoperators: $\mathcal{L}_{\s + \PM}$ and  $\mathcal{L}_{\s + \text{ADO}}$. (d) The non-Markovian EPs can then be identified by observing the complex spectrum of these extended Liouvillian superoperators. }
\label{fig: illust}
\end{figure*}

In this section, we present a general approach to characterize EPs for open systems subject to non-Markovian noise. To begin with, we consider an open quantum system (S) coupled to a bosonic environment (E), where the total Hamiltonian is expressed by 
\begin{equation}
    \begin{aligned}
        &H_\tot =H_\s + H_\e + H_{\s\e} \text{~with~~} H_\e = \sum_k \omega_k b^\dag_k b_k, \\
        &H_{\s\e}= QX,~\text{and } X = \sum_k g_k  (b^\dag_k +  b_k). \label{eq: H_tot}
    \end{aligned}
\end{equation}
Here, $H_\s$ and $H_\e$ represent the free Hamiltonians of the system and its environment, while $H_{\s\e}$ describes their interactions. Also, $\omega_k$ and $g_k$ correspond to the frequency and the coupling strength for the environmental mode $b_k$, respectively, and $Q$ represents an arbitrary system-environment coupling operator. We consider that the environment is initialized in a Gibbs state at temperature $T$: $\rho_\e = \exp(-\beta H_\e)/\tr[\exp(-\beta H_\e)]$, where $\beta=(k_\text{B} T)^{-1} $ with $k_\text{B}$ denoting the Boltzmann constant. In this case, the exact dynamics of the open system's reduced density matrix (within the interaction picture) can be written as 
\begin{equation}
    \begin{aligned}
        \rho_\s(t) = \hat{\mathcal{T}}\exp\left\{\hat{\mathcal{F}}\Big[Q,C(t)\Big]\right\} \rho_\s(0).\label{eq: exact dynamics}
    \end{aligned}
\end{equation}
Here, $\hat{\mathcal{T}}$ denotes the time-ordering operator and $\hat{\mathcal{F}}$ represents the Feynman-Vernon influence functional~\cite{breuer2002theory}, which can be expressed by
\begin{equation}
    \begin{aligned}
        &\hat{\mathcal{F}} = -\int_{0}^{t} dt_1 \int_{0}^{t_1} dt_2~ Q(t_1) ^\times [C_{\mathbb{R}}(t_1 - t_2) Q(t_2)^\times \\
        &~~~~~~~~~~~~~~~~~~~~~~~~~~~~~~~~~~~+ i C_{\mathbb{I}}(t_1 - t_2) Q(t_2)^\circ ],
    \end{aligned}
\end{equation}
where $Q(t) = \exp(iH_\s t)Q\exp(-iH_\s t)$, and we adopt the superoperator notations: 
\begin{equation}
    \begin{aligned}
        Q(t)^\circ = \{Q(t), \bullet\} 
        \text{~and~}Q(t)^\times = [Q(t), \bullet].
    \end{aligned}
\end{equation}
In addition, $C_{\mathbb{R}(\mathbb{I})}$ denotes the real (imaginary) part of the environmental correlation function 
\begin{equation}
    C(t)=\tr[X(t) X(0) \rho_\e]
\end{equation}
with $X(t) = \exp(iH_\e t)X\exp(-iH_\e t)$.
An essential feature of $\hat{\mathcal{F}}$ is its exclusive dependence on the system-environment coupling operator $Q$ and the environmental correlation function $C(t)$. The latter can be expressed by
\begin{equation}
    C(t)=\int_0^\infty d\omega~ \frac{J(\omega)}{\pi}\left[\coth\left(\frac{\beta \omega}{2}\right)\cos(\omega t) - i\sin(\omega t)\right], \label{eq: correlation}
\end{equation}
where $J(\omega)=\sum_k |g_k|^2 \delta(\omega-\omega_k)$ represents the coupling spectral density. This property enables the reproduction of the exact open system dynamics through an auxiliary model involving a small set of fictitious damping modes, i.e., the pseudomodes, provided that the correlation function for the artificial model aligns with Eq.~\eqref{eq: correlation}. 

For a broad range of cases, the correlation function can be efficiently expressed as a finite weighted summation of exponential terms, i.e., 
\begin{equation}
    C(t) = \sum_i \alpha_i^2 \exp(-i\Omega_i t -\gamma_i |t|/2 ). \label{eq: correlation_exponential_decomposition}
\end{equation}
With this expression, one can construct the PMEOM:
\begin{equation}
    \begin{aligned}
        &\frac{d}{dt}\rho_{\s+\PM}(t)= \mathcal{L}_{\s+\PM}[\rho_{\s+\PM}(t)]\\
        &=-i[H_{\s+\PM},\rho_{\s+\PM}(t)]+\sum_i \gamma_i \mathcal{L}_{a_i}[\rho_{\s+\PM}(t)], \\
        &\text{with~~}H_{\s+\PM} = H_\s +\sum_i \Omega_i a_i^\dag a_i + \alpha_i Q( a^\dag_i + a_i), \label{eq: PMEOM}
    \end{aligned}
\end{equation}
where, $\{a_i\}$ represent the pseudomodes, and we introduce the dissipator $\mathcal{L}_{a_i}[{\bullet}]=a_i \bullet a_{i}^\dag-\{a_i^\dag a_i,\bullet\}/2$. Notably, the environmental influences on the open system are now captured by the pseudomodes' frequencies $\Omega_i$, damping rates $\gamma_i$, and the system-pseudomode coupling strengths $\alpha_i$. The exact dynamics of S can be obtained by tracing out the pseudomodes (PM), i.e., $\rho_\s(t) = \tr_\PM[\rho_{\s+\PM}(t)]$, after solving the PMEOM with these pseudomodes initialized in the thermal states.   

One notable benefit of employing the pseudomode model lies in its facilitation of establishing physical intuitions regarding the EP criteria. For instance, recent works have suggested that EPs are closely related to critical damping points for both classical and quantum systems~\cite{PRXQuantum.2.040346}. As we demonstrate below, it is feasible to pinpoint EPs by balancing the system-pseudomode coupling strength and the pseudomode damping, leading the system to a critical damping point. 

Furthermore, a systematic procedure for characterizing non-Markovian EPs based on a standard spectral analysis can be established. The main idea is grounded in the observation that both the temporal evolution of $\rho_{\s+\PM}(t)$ and $\rho_{\s}(t)$ are governed by the spectral properties of the extended Liouvillian superoperator $\mathcal{L}_{\s+\PM}$. Specifically, assuming that $\mathcal{L}_{\s+\PM}$ is diagonalizable, we consider its eigenvalues and the corresponding eigenmatrices: $\{\lambda_i,\hat{\rho}_{\s+\PM,i}\}_i$. The dynamics of $\rho_{\s+\PM}(t)$ can then be expressed by 
\begin{equation}
    \rho_{\s+\PM}(t) = \sum_i c_i \exp(\lambda_i t) \hat{\rho}_{\s+\PM, i}.
\end{equation} 
By tracing out these pseudomodes, one can observe that the exact dynamics of the system reduced state follows a similar expression, replacing these eigenmatrices $\hat{\rho}_{\s+\PM, i}$ with the reduced eigenmatrices $\hat{\rho}_{\s, i}=\tr_{\PM}(\hat{\rho}_{\s+\PM, i})$, namely, 
\begin{equation}
    \rho_{\s}(t) = \sum_i c_i \exp(\lambda_i t) \hat{\rho}_{\s, i}.
\end{equation}

To describe EPs, one considers a family of parametrized extended Liouvillian superoperators $\mathcal{L}_{\s+\PM}(\boldsymbol{\xi})$, bearing in mind that $\boldsymbol{\xi}$ includes the parameters related to both the system and the structured environments. Since $\mathcal{L}_{\s+\PM}(\boldsymbol{\xi})$ is generally non-Hermitian, EPs could potentially exist in the parameter space. For instance, let $\boldsymbol{\xi}_{\ep n}$ represent an $n$th-order EP in the parameter space, where $n$ different eigenvalues and the corresponding eigenmatrices $\{\lambda_i,\hat{\rho}_{\s+\PM,i}\}_{i\in A}$
coalesce into $\{\lambda_{\ep}, \hat{\rho}_{\s+\PM,\lambda_{\ep}}\}$. Here, $A$ denotes a set of indices. Due to the coalescence of the eigenmatrices, the corresponding $n$-dimensional eigensubspace for $\mathcal{L}_{\s+\PM}(\boldsymbol{\xi}_\ep)$ cannot be diagonalized. Nevertheless, a Jordan block for the subspace can be constructed by introducing generalized eigenmatrices $\{\hat{\rho}^{(j)}_{\s+\PM,\lambda_{\ep}}\}_{j=0,\cdots,n-1}$~\cite{moiseyev2011non}, such that the system dynamics can be expressed by
\begin{equation}
    \begin{aligned}
        \rho_{\s}(t) = \sum_{i\notin A} c_i e^{\lambda_i t} \hat{\rho}_{\s,i} 
        +e^{\lambda_{\ep} t}\sum_{j=0}^{n-1}\sum_{m=0}^{j} \frac{t^m \tilde{c}_m}{m!}\hat{\rho}^{(j)}_{\s,\lambda_{\ep}},
    \end{aligned}
    \label{eq: rhos_spectral}
\end{equation}
where the reduced generalized eigenmatrices are introduced as
\begin{equation}
    \hat{\rho}^{(j)}_{\s,\lambda_{\ep}}=\tr_{\PM}(\hat{\rho}^{(j)}_{\s+\PM,\lambda_{\ep}}).
\end{equation} 
Equation~\eqref{eq: rhos_spectral} suggests that the polynomial time dependence, a common dynamical signature of EPs, could be observed in the reduced dynamics of the open system. In essence, the PMEOM provides an intuitive and direct route to investigate EPs beyond the Markovian limit, and this approach is compatible with the conventional spectral analysis by introducing the (generalized) reduced eigenmatrices.

\subsubsection{Hierarchical equations of motion}
It is worth noting that an alternative approach is to consider the framework of HEOM. In this context, a set of auxiliary density operators (ADOs) is introduced to capture the non-Markovian and non-perturbative effects~\cite{Tanimura01,Huang2023, Mauro2022,KondoQED2023,Tanimura02,Tanimura05}. Similarly, we can define the extended quantum state $\rho_{\s+\text{ADO}}$ that contains both the system reduced state and the ADOs. The dynamics of the extended state is governed by the extended Liouvillian superoperator $\mathcal{L}_{\s+\text{ADO}}$. The system reduced state can be obtained through a linear operation, specifically $\rho_\s(t)=\mathcal{P}[\rho_{\s+\text{ADO}}(t)]$, where $\mathcal{P}$ is a superoperator for discarding all ADOs (see Methods). Therefore, the EPs for non-Markovian open quantum systems can also be equivalently characterized under the framework of the HEOM by introducing the corresponding (generalized) reduced eigenmatrices, which are expressed as $\tilde{\rho}_{\s,i}=\mathcal{P}(\tilde{\rho}_{\s+\text{ADO},i})$ and $\tilde{\rho}^{(j)}_{\s,\lambda_\ep}=\mathcal{P}(\tilde{\rho}^{(j)}_{\s+\text{ADO},\lambda_\ep})$.

\subsection*{Spin-boson model}
\begin{figure*}
\includegraphics[width=1\linewidth]{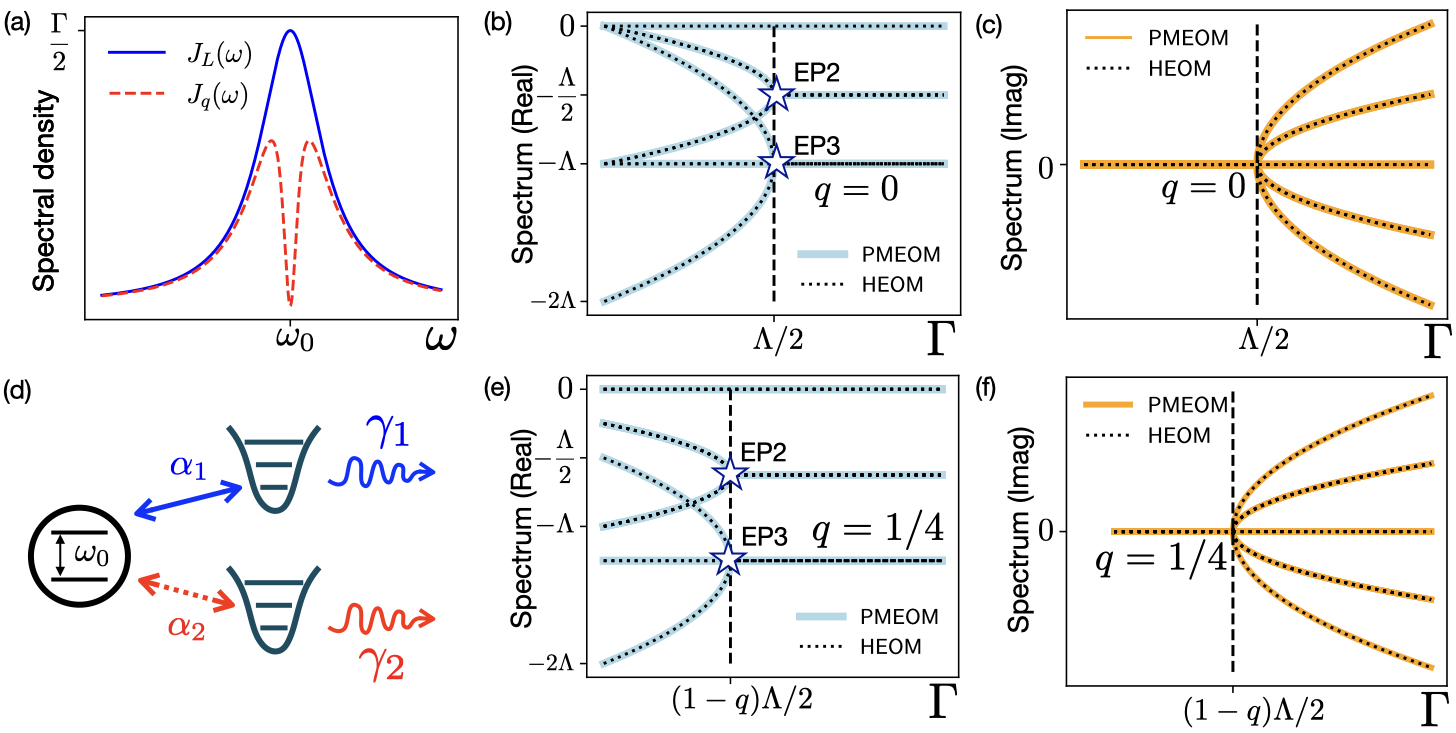}
\caption{(a) Lorentzian $J_L(\omega)$ and band gap $J_q(\omega)$ spectral densities centered at the qubit transition frequency $\omega_0$. (d) The effects of these spectral densities can be represented by two pseudomodes: The Lorentzian spectral density can be described by the upper mode with qubit-mode coupling strength $\alpha_1=\sqrt{\Gamma \Lambda/2}$ and mode's damping rate $\gamma_1 =2\Lambda$, while the band gap is characterized by the lower mode with the non-Hermitian coupling $\alpha_2=i\sqrt{q\Gamma \Lambda/2}$ and damping rate $\gamma_2=2 q \Lambda$, where the red dashed line signifies the unphysical nature of this mode. (b, c) The real and imaginary parts of the spectrum of the extended Liouvillian for the gapless scenario ($q=0$). Two exceptional points, EP2 and an EP3 emerge at the coupling strength $\Gamma=\Lambda/2$. (e, f) The real and imaginary parts of the spectrum of the extended Liouvillian when $q=1/4$. The EP criterion becomes $\Gamma=(1-q)\Lambda/2$. The dotted curves in (b, c, e, f) represent the spectrum of the extended superoperators for HEOM (see Supplementary Note 1 for detailed derivations).}
\label{fig: spin-boson spectrum}
\end{figure*}

To gain a more intuitive understanding, we apply the PMEOM to the prototype spin-boson model, showing that a pure non-Markovian EP can be observed by adjusting the structural characteristics of the environment's spectral density. To simplify our analysis, we consider a zero-temperature environment and adopt the rotating-wave approximation (RWA), where the system-environment interaction in Eq.~\eqref{eq: H_tot} and the system-pseudomode coupling in Eq.~\eqref{eq: PMEOM} can be expressed by $\sum_{k}g_k (\tilde{Q} b_k^\dag + \tilde{Q}^\dag b_k)$ and $\sum_{i}\alpha_i (\tilde{Q} a_i^\dag + \tilde{Q}^\dag a_i)$, respectively. Furthermore, we consider a scenario where the spectral density function $J(\omega)$ is well localized in the vicinity of a high frequency $\tilde{\omega}\gg 0$, enabling us to effectively approximate the correlation in Eq.~\eqref{eq: correlation} by extending the lower limit of the integral to negative infinity. Note that the assumptions mentioned above are also considered in the original proposals of PMEOM~\cite{PhysRevA.55.2290, PhysRevA.64.053813, PhysRevA.80.012104}.

The model involves a qubit representing the open system, where the system Hamiltonian and system-environment coupling operator are $H_\s=\omega_0 |e\rangle \langle e|$ and $\tilde{Q}=\sigma_-$, respectively. Here, $\omega_0$ denotes the qubit transition frequency between the ground state $|g\rangle$ and the excited state $|e\rangle$, and $\sigma_+=|e\rangle \langle g|$ ($\sigma_-=|g\rangle \langle e|$) represents the raising (lowering) operator. We consider a Lorentzian spectral density that is expressed by
\begin{equation}
    J_L(\omega) = \frac{1}{2}\frac{\Gamma \Lambda^2}{(\omega-\omega_0)^2 +\Lambda^2}, \label{eq: J_L}
\end{equation}
where $\Gamma$ and $\Lambda$ denote the coupling strength and the spectral width, respectively. In the interaction picture and under the earlier specified assumption of extending the spectral density function to negative frequencies, the environmental correlation function can be expressed by a single exponential term, i.e., $C(t) = (\Gamma \Lambda/2) \exp(-\Lambda |t|)$. Therefore, the PMEOM can be constructed by introducing a single pseudomode with the damping rate $\gamma = 2\Lambda$ and the qubit-pseudomode coupling strength $\alpha = \sqrt{\Gamma \Lambda/2}$. As aforementioned, the pseudomode representation provides a physical intuition that the EP could be located by balancing $\gamma$ and $\alpha$ (or $\Gamma$ and $\Lambda$) so that the system reaches a critical damping point. 

The corresponding extended Liouvillian superpoerator $\mathcal{L}_{\s+\PM}(\Gamma,\Lambda)$ can be described by a $9\times 9$ non-Hermitian matrix in the single-excitation subspace (Supplementary Note 1). The spectrum is illustrated in Fig.~\ref{fig: spin-boson spectrum}, revealing that $\Gamma=\Lambda/2$ corresponds to a second-order EP (EP2) and a third-order EP (EP3). Notably, these EPs are purely non-Markovian, because they are unobservable in the Markovian limit. Specifically, in such a limit, the spectral width (and thus the damping rate of the pseudomode) becomes infinite, $\Lambda \rightarrow \infty$. Therefore, the pseudomode can be adiabatically eliminated and the dynamics is governed by a qubit-only Markovian master equation, i.e., $\dot{\rho}_\s(t) = \Gamma[2\sigma_-\rho_\s(t) \sigma_+ -\{\sigma_+ \sigma_-,\rho_\s(t)\}]/2$. Intuitively, there is only one qubit decay channel without internal tunneling between the qubit energy levels, thereby EP does not emerge in this scenario~\cite{PhysRevA.100.062131}. 

Let us proceed with a deeper analysis of the EP dynamical signatures. According to Figs.~\ref{fig: spin-boson spectrum}(b) and \ref{fig: spin-boson spectrum}(c), the EP condition pinpoints a real-to-complex transition in the spectrum. In other words, the EP condition corresponds to a critical damping point, separating the overdamped (pure decay) and underdamped (oscillatory) regimes. By investigating the generalized eigenmatrices for the extended superoperator at the EP condition, i.e., $\mathcal{L}_{\s+\PM}(\Gamma=\Lambda/2)$, we conclude that the qubit coherence and population dynamics are respectively determined by the EP2 and EP3, resulting in observing first-order and second-order polynomial time dependencies:
\begin{equation}
    \begin{aligned}
        &\langle e|\rho_\s(t)| g \rangle = \frac{1}{2} (\Lambda  t+2) e^{-\frac{1}{2} \Lambda  t}\langle e|\rho_\s(0)| g \rangle, \\
        &\langle e|\rho_\s(t)| e \rangle = \frac{1}{4} (\Lambda^2 t^2+4 \Lambda t + 4) e^{-\Lambda  t}\langle e|\rho_\s(0)| e \rangle.
    \end{aligned}
\end{equation}

Notably, by utilizing the Breuer-Lain-Pillo~\cite{PhysRevLett.103.210401} and the Riva-Huelga-Plenio~\cite{PhysRevLett.105.050403} non-Markovian measures to the qubit dynamics, we find that the EP condition also aligns with the onset of non-Markovianity (see Supplementary Note 2). This reveals an unexpected connection between the Markovian-to-non-Markovian and non-Hermitian EP phase transitions.

Furthermore, the transition from the overdamped to underdamped regimes opens up the possibility of observing enhanced sensitivity to external perturbations in the vicinity of the EP. For instance, consider a small perturbation in the coupling strength, represented as $\Gamma\rightarrow \Lambda(1+\epsilon)/2$, with $\epsilon>0$. This perturbation induces splittings in the imaginary parts of the eigenvalues, resulting in oscillatory dynamics. Notably, these splittings scale as $\sqrt{\epsilon}$, indicating the sensitivity enhancement in the characteristic frequency of the oscillatory behavior. For this model, both the qubit excited state population and coherence vanish periodically. We choose, for simplicity, the first vanishing time $t_{\text{vanish}}$ of the qubit coherence to capture the system's oscillations. One can find $t_{\text{vanish}}^{-1}\approx\Lambda  \sqrt{\epsilon }/2 +O\left(\epsilon\right)$, which implies that $t_{\text{vanish}}$ is sensitive to the external perturbation when the system is prepared at the EP.

Up to this point, we only tune the environmental parameters without changing the shape of the Lorentzian profile, and we have shown that such an adjustment does not modify the structure of the PMEOM. Let us further consider a more complex scenario, where an additional parameter can drastically modify the spectral shape. To this end, we introduce a band-gap structure to the Lorentzian background~\cite{breuer2002theory}, which is modeled as
\begin{equation}
    \begin{aligned}
        J_q(\omega) = J_L(\omega)-\frac{1}{2}\frac{\Gamma (q\Lambda)^2}{(\omega-\omega_0)^2 +(q\Lambda)^2},
    \end{aligned}
\end{equation}
where the band gap is located at the frequency $\omega_0$, such that $J_q(\omega_0)=0$~\cite{breuer2002theory} [see also Fig.~\ref{fig: spin-boson spectrum}(a)], and the parameter $q \in (0,1]$ is used to control the relative width for the band gap. The environmental correlation function is now expressed by two exponential terms:
\begin{equation}
    C(t) = \frac{\Lambda\Gamma}{2} \exp(-\Lambda |t|)-\frac{q\Lambda\Gamma}{2} \exp(-q\Lambda |t|).
\end{equation}
Accordingly, the PMEOM for the exact dynamics involves two pseudomodes ($a_1$ and $a_2$), each characterized by the qubit-pseudomode coupling strengths and pseudomode damping rates [see also Fig.~\ref{fig: spin-boson spectrum}(d)]:
\begin{equation}
    \{\alpha_1 = \sqrt{\Gamma \Lambda/2}, \alpha_2=i\sqrt{q}\alpha_1,\gamma_1 = 2\Lambda, \gamma_2=q\gamma_1\}.
\end{equation}
The spectrum of the corresponding extended Liouvillian superoperator is presented in Figs.~\ref{fig: spin-boson spectrum}(e) and ~\ref{fig: spin-boson spectrum}(f), suggesting that the EP criterion becomes 
\begin{equation}
    \Gamma=(1-q)\Lambda/2.
\end{equation}
Consequently, including a band gap into the environmental spectral profile leads to a displacement of the EP in the parameter space. In this case, the system requires a smaller coupling strength to reach the EP criterion compared to the case of the previous gapless Lorentzian environment. 

We remark that $\alpha_2$ is purely imaginary, so it does not directly correspond to a physical mode.  This characteristic causes the $H_{\s+\PM}$ to become non-Hermitian~\cite{Lambert2019}. Nevertheless, the reduced dynamics of the qubit is still exact and well-behaved as long as the PMEOM is consistent with the given $C(t)$~\cite{Lambert2019}. Furthermore, in Fig.~\ref{fig: spin-boson spectrum} and Supplementary Note 1, we demonstrate that the spectrum of the extended Liouvillian operators for the HEOM is consistent with that of PMOEM.

\subsection*{Linear bosonic systems with adjoint pseudomode-equation of motion}
The PMEOM in Eq.~\eqref{eq: PMEOM} offers significant flexibility by allowing both the system Hamiltonian $H_\s$ and the coupling operator $Q$ to be arbitrary. This utility facilitates the extension of our analysis to a variety of open quantum systems. As an illustrative application, we now explore linear bosonic systems, which have garnered significant interest in the study of both LEPs and HEPs~\cite{arkhipov2023dynamically,
PhysRevA.101.013812,
PhysRevA.102.033715,
PhysRevA.104.012205,
perina2022quantum}. Specifically, we examine a general Hamiltonian with $M$ coupled modes, expressed as 
\begin{equation}
    H_\s = \sum_{k=1}^M \Omega_k c^\dag_k c_k + \sum_{j<k}\chi_{j,k}(c^\dag_j c_k + c_k^\dag c_j),
\end{equation}
where $\Omega_k$ denotes the frequency associated with the mode $c_k$, and $\chi_{j,k}$ represents the coherent coupling between the modes $j$ and $k$. In addition, we consider that the $M$th mode further couples to the environment with the coupling operator $Q=c_M$. In the Heisenberg picture, the dynamics is governed by the following adjoint PMEOM:
\begin{equation}
    \begin{aligned}
        &\frac{d}{dt}O_{\s+\PM}(t)=i[H_{\s+\PM},O_{\s+\PM}(t)]\\
        &+\sum_i \frac{\gamma_i}{2} (2 a^\dag_i O_{\s+\PM}(t) a_i -\{a_i^\dag a_i,O_{\s+\PM}(t)\}).
    \end{aligned}
\end{equation}
To facilitate the analysis, we define a vector containing the annihilation operators for both the system modes and the pseudomodes: 
\begin{equation}
    \mathbf{v}=\big(c_1,\cdots,c_M,a_1,\cdots, a_i,\cdots\big)^T.
\end{equation}
Focusing on the amplitudes of these modes, we denote 
\begin{equation}
   \begin{aligned}
        &\langle\mathbf{v(t)}\rangle = (\langle c_1(t)\rangle,\cdots,\langle c_M(t)\rangle,\langle a_1(t)\rangle,\cdots, \langle a_i(t)\rangle,\cdots)^T, \\
        &\text{with~~} \langle a(t) \rangle =\tr[a(t) \rho_{\s+\PM}(0)].
   \end{aligned}
\end{equation}
Through the adjoint PMEOM, the dynamics for the amplitudes is determined by the effective NHH
\begin{equation}
    H_{\text{eff},\s+\PM} = H_{\s+\PM}-i\sum_i\gamma_{i}a^\dag_i a_i/2 = \mathbf{v}^\dag \mathbf{H}_{\text{eff},\s+\PM}\mathbf{v},
\end{equation}
where $\mathbf{H}_{\text{eff},\s+\PM}$ is a matrix representation of the NHH. Specifically, one obtains 
\begin{equation}
    \langle\dot{\mathbf{v}}(t)\rangle=i \mathbf{H}_{\text{eff},\s+\PM}\langle\mathbf{v}(t)\rangle,
\end{equation}
indicating that potential EPs can now be encoded in $\mathbf{H}_{\text{eff},\s+\PM}$. 

\subsubsection{Example: two-mode system coupled to a Lorentzian environment}
For a concrete example, we examine a two-mode system coupled to a Lorentzian environment, as described in Eq.~\eqref{eq: J_L}, assuming that 
\begin{equation}
    \chi_{1,2}=\chi \text{~and~}\Omega_1=\Omega_2=\omega_0.
\end{equation}
In the Markovian limit ($\Lambda\rightarrow \infty$), the resulting system-only effective non-Hermitian Hamiltonian within the rotating frame is given by:
\begin{equation}
    \mathbf{H}_{\text{eff},\s}=\begin{pmatrix}
        0&\chi\\
        \chi&i\Gamma
        \end{pmatrix}.
\end{equation}
The corresponding eigenvalues are $ (i \Gamma \pm \sqrt{4 \chi ^2-\Gamma ^2} )/2$, indicating the presence of an EP2 if $|\chi|=\Gamma/2$, with the degenerate eigenvalue $i\Gamma/2$. With a finite width $\Lambda$, the effective Hamiltonian takes the form:
\begin{equation}
    \mathbf{H}_{\text{eff},\s+\PM}=\begin{pmatrix}
        0   & \chi & 0\\
        \chi& 0 &\sqrt{\frac{\Gamma\Lambda}{2}}\\
        0 & \sqrt{\frac{\Gamma\Lambda}{2}} & i \Lambda
        \end{pmatrix}.
\end{equation}
By matching the coefficients of the characteristic polynomial, an EP3 is identified with the following criteria: 
\begin{equation}
    \left\{|\chi| = \frac{\Lambda}{3\sqrt{3}},~\Gamma=\frac{16\Lambda}{27}\right\},
\end{equation}
and the degenerate eigenvalue is $i\Lambda/3$. In other words, the EP can be transformed from second to third order with the introduction of the structured environmental characteristics. 

This upgrade can lead to a further enhancement in the system sensitivity. For instance, we introduce a perturbation $\epsilon>0$ to the coupling strength $\chi\rightarrow \chi(1+\epsilon)$. For the scenario in the exact Markovian limit, the eigenvalues in the vicinity of the EP2 are 
\begin{equation}
    \left\{i \frac{\Gamma}{2}-\frac{\Gamma  \sqrt{\epsilon }}{\sqrt{2}}+O\left(\epsilon ^{\frac{3}{2}}\right),i \frac{\Gamma}{2}+\frac{\Gamma  \sqrt{\epsilon }}{\sqrt{2}}+O\left(\epsilon ^{\frac{3}{2}}\right)\right\}.
\end{equation}

In contrast to this case, for the scenario with a finite spectral width, the eigenvalues in the vicinity of the EP3 take the form 
\begin{equation}
    \begin{aligned}
        \Big\{&i\frac{\Lambda}{3}+ x_1\Lambda \epsilon^{\frac{1}{3}}+O(\epsilon^{\frac{2}{3}}),i\frac{\Lambda}{3}+ x_2\Lambda \epsilon^{\frac{1}{3}}+O(\epsilon^{\frac{2}{3}}),\\
        &i\frac{\Lambda}{3}+ x_3\Lambda \epsilon^{\frac{1}{3}}+O(\epsilon^{\frac{2}{3}})\Big\},
    \end{aligned}
\end{equation}
where $x_1$, $x_2$, and $x_3$ are constants. We observe a change from a square-root bifurcation for the Markovian EP2 to a cubic-root bifurcation for the non-Markovian EP3 in response to the external perturbation, signifying the enhancement of the system sensitivity to the external perturbation. 

Intuitively, one may anticipate that higher-order EPs may emerge when considering an environment with a more complicated spectral structure, which requires more pseudomodes to capture the non-Markovian dynamics and increases the dimension of the non-Markovian NHHs

\section{discussion}
We have presented a general theory on characterizing non-Markovian EP based on pseudomode mapping and hierarchical equations of motion. This approach is applied to both the spin-boson model and linear bosonic systems, uncovering the presence of a purely non-Markovian EP that corresponds to the transition from Markovian to non-Markovian dynamics. We demonstrate that adjustments to the EP criteria can be accomplished through the modification of the environmental spectral profile. Additionally, the incorporation of non-Markovian effects can increase the order of the EP by effectively increasing the dimension of both the extended Liouvillian superoperator and NHHs. This presents an innovative strategy for identifying higher-order EPs. 

A direct extension involves exploring more realistic examples that do not rely on the RWA and restore the detailed-balance condition~\cite{Lambert2019}. In addition, although this work focuses exclusively on a bosonic environment, the proposed framework can be directly generalized to the scenarios with arbitrary combinations of bosonic and fermionic baths~\cite{Tanimura01,
Mauro2022,
lambert2020bofinheom,
Huang2023}. Moreover, beyond the PMEOM and HEOM, our method of describing non-Markovian EPs via extended Liouvillian superoperators can also be applied to other pertinent methodologies, such as the dissipaton-embedded master equation~\cite{yan2014theory,
wang2022quantum} and reaction-coordinate mapping~\cite{PhysRevA.90.032114,
iles2016energy,
PhysRevB.100.035109,
PhysRevA.104.052617}.

Future work involves further generalizing the theory of non-Markovian EPs. An intriguing possibility involves extending the hybrid-Liouvillian formalism~\cite{PhysRevA.101.062112} to the non-Markovian domain by incorporating the postselection of quantum trajectories. 
Additionally, it is worthwhile to delve into the potential applications emerging from the intricate interplay between the (non-)Markovian exceptional and diabolic points~\cite{arkhipov2023dynamically,perina2022quantum} or the exotic topology and geometry of the parameter space~\cite{ding2022non,
RevModPhys.93.015005,
PhysRevResearch.4.023070,
ju2024emergent,ju2024event}. Such investigations could uncover new aspects of non-Markovian EPs, enhancing our understanding of open quantum systems embedded in environments with memory effects.
\section{Methods}
\subsection*{The extended Liouvillian superoperators for HEOM}
Here, we introduce the extended Liouvillian superoperators within the HEOM formalism, considering both scenarios with and without the rotating wave approximation (RWA). For the case without RWA, we decompose the correlation function into exponential terms [as in Eq.~\eqref{eq: correlation_exponential_decomposition}]: 
\begin{equation}
    C(\tau)=\sum_{u=\mathbb{R},\mathbb{I}}(\delta_{u,\mathbb{R}}+i\delta_{u,\mathbb{I}})C^{u}(\tau)
\end{equation}
with $\mathbb{R}$ and $\mathbb{I}$ representing real and imaginary parts, and 
\begin{equation}
    C^{u}(\tau)=\sum_{l}^{l_{\text{max}}}\xi_{l}^{u}\exp(-\chi_{l}^{u}\tau).
\end{equation}
Through iterative time differentiation of the exact dynamics in Eq.~\eqref{eq: exact dynamics}, the HEOM can be expressed as a time-local differential equation within an expanded space formed by the auxiliary density operators (ADOs):~\cite{Huang2023, Mauro2022,KondoQED2023,Tanimura03,Tanimura04}
\begin{equation}
\begin{aligned}
\frac{d}{dt}\rho_{\text{S}+\text{ADO}}(t)=\mathcal{L}_{\text{S}+\text{ADO}}\left[\rho_{\text{S}+\text{ADO}}(t)\right].
\label{eq:HEOML1}
\end{aligned}
\end{equation}
Here, we label the ADOs with a vector $\bfj=[j_m,\cdots,j_1]$, linking each ADO to a specific exponential term in the correlation function. The system reduced density operator and these ADOs can then be expressed as $\rho^{(m)}_{\bfj}(t)$, such that
\begin{equation}
    \rho_{\s+\ADO}=
\begin{bmatrix}
\rho^{(0)}_{\bfj }(t) \\
\rho^{(m>0)}_{\bfj }(t) \\
\vdots
\end{bmatrix},
\end{equation}
where $m$ indicates the hierarchical level of the ADO with $\rho_\bfj^{(0)}(t)=\rho_\s(t)$. Consequently, the system reduced state can be obtained by performing a projector $\mathcal{P}$ on the $\rho_{\s+\ADO}$:
\begin{equation}
\begin{aligned}
\begin{bmatrix}
\rho_{\text{S}}(t) \\
0 \\
\vdots  \\
0
\end{bmatrix}
 = \mathcal{P}[\rho_{\text{S}+\text{ADO}}(t)]=
\begin{bmatrix}
\mathbb{1} & \dots & 0\\
\vdots & \ddots & \vdots\\
 0 & \dots & 0
\end{bmatrix}
\begin{bmatrix}
    \rho^{(0)}_{\bfj }(t) \\
    \rho^{(m>0)}_{\bfj }(t) \\
    \vdots
\end{bmatrix},
\label{eq:rhoS}
\end{aligned}
\end{equation}%
where $\mathbb{1}$ is an identity matrix with the same dimension as the system reduced density operator. Under this expression of ADOs, the corresponding extended Liouvillian superoperator of HEOM can be written as~\cite{Mauro2022,Huang2023}

\begin{equation}
    \begin{aligned}
        &\mathcal{L}_{\text{S}+\text{ADO}}[\rho^{(m)}_{\bfj}(t)]
        =\mathcal{L}_{0}[\rho^{(m)}_{\bfj}(t)]
        -\sum_{r=1}^{m}\chi_{j_{r}}\rho^{(m)}_{\bfj}(t)\\
        &~~~~-i\sum_{j'} \mathcal{A}_{j'}
        [\rho^{(m+1)}_{\bfj^+}(t)]
        -i\sum_{r=1}^{m}\mathcal{B}_{j_{r}}
        [\rho^{(m-1)}_{\bfj_{r}^{-}}(t)],
        \label{eq:HEOML2}
    \end{aligned}
\end{equation}
where a multi-index notation is used: $\bfj^{+}=[j',j_{m},\cdots,j_1]$, and $\bfj_{r}^{-}=[j_m,\cdots,j_{r+1},j_{r-1},\cdots,j_{1}]$, and $\mathcal{L}_{0}[\cdot]=-iH_{\text{S}}^{\times}$. The system-environment interaction is encoded in the superoperators $\hat{\mathcal{A}}_{j}$ and $\hat{\mathcal{B}}_{j}$ that couple the $m$th-level bosonic ADOs to the $(m+1)$th- and $(m-1)$th-levels, respectively. Their explicit expressions are given by
\begin{equation}
\begin{aligned}
\mathcal{A}_{j}[\cdot] = Q^{\times} \text{~and~} \mathcal{B}_{j}[\cdot] = \delta_{u,\mathbb{R}}~\xi^{\mathbb{R}}_{l}~Q^{\times}+i\delta_{u,\mathbb{I}}~\xi^{\mathbb{I}}_{l}~Q^{\circ}.
\label{eq:A}
\end{aligned}
\end{equation}
Let us now consider the case with RWA, where the system-environment interaction is written as $H_{\text{SE}} = \sum_{k}g_k (\tilde{Q} b_k^\dag + \tilde{Q}^\dag b_k).$
In this case, we separate the correlation function into the absorption $(\nu = +)$ and emission $(\nu =-)$ components, namely $C(t)
=\sum_{\nu=\pm}C^{\nu}(t)$, where
\begin{equation}
    \begin{aligned}
        &C^{+}(t)
        =\frac{1}{\pi}\int_{0}^{\infty} d\omega 
        J(\omega)n^{\textrm{eq}}(\omega)e^{i\omega \tau}\\       &~~~~~~~~=\sum_{l}^{l_{\text{max}}}\xi_{l}^{\nu=+}\exp(-\chi_{l}^{\nu=+}\tau),\\
        \text{and~}&C^{-}(t)=\frac{1}{\pi}\int_{0}^{\infty} d\omega 
J(\omega)[n^{\textrm{eq}}(\omega)+1]e^{-i\omega \tau}\\
&~~~~~~~~=\sum_{l}^{l_{\text{max}}}\xi_{l}^{\nu=-}\exp(-\chi_{l}^{\nu=-}\tau).
        \label{eq:C_b}
    \end{aligned}
\end{equation}
Here $n^{\textrm{eq}}(\omega)=\{\exp[\omega/k_{\textrm{B}}T]-1\}^{-1}$ represents the Bose–Einstein distribution with $k_{\textrm{B}}$ as the Boltzmann constant and $T$ as the temperature, and $J(\omega)$ represents the spectral density of the environment. 

The Feynman-Vernon influence functional can be derived~\cite{Mauro2022,Huang2023}:
\begin{equation}
    \begin{aligned}
        &\hat{\mathcal{F}}[\cdot] = -\sum_{\nu=\pm} \int_{0}^{t} dt_1 \int_{0}^{t_1} dt_2~ \tilde{Q}^{\bar{\nu}}(t_1)^{\times} 
        \\&~~~ \Big\{C^{\nu}(t_1 - t_2) \tilde{Q}^{\nu}(t_2)[\cdot] + C^{\bar{\nu}\ast}(t_1 - t_2) [\cdot] \tilde{Q}^{\nu}(t_2)\Big\},
    \end{aligned}
\end{equation}
where $\tilde{Q}^{\nu=+}(t)=\tilde{Q}^{\dagger}(t)$ and $\tilde{Q}^{\nu=-}(t)=\tilde{Q}(t)$. The corresponding extended Liouvillian superoperator can then be expressed as
\begin{equation}
    \begin{aligned}
        \mathcal{L}_{\text{S}+\text{ADO}}^{\text{RWA}}[\rho^{(m)}_{\bfk}(t)]
        =\mathcal{L}_{0}[\rho^{(m)}_{\bfk}(t)]
        -\sum_{r=1}^{m}\chi_{k_{r}}\rho^{(m)}_{\bfk}(t)
        \\-i\sum_{k'} \mathcal{A}_{k'}
        [\rho^{(m+1)}_{\bfk^+}(t)]
        -i\sum_{r=1}^{m}\mathcal{B}_{k_{r}}
        [\rho^{(m-1)}_{\bfk_{r}^{-}}(t)],
        \label{eq:HEOML_RWA}
        \end{aligned}
    \end{equation}%
where the superoperators $\mathcal{A}_{k'}$ and $\mathcal{B}_{k_{r}}$ are modified as follows
\begin{equation}
\begin{aligned}
\mathcal{A}_{k}[\cdot] = \tilde{Q}^{\bar{\nu}\times}~\text{and~} \mathcal{B}_{k}[\cdot] = \xi^{\nu}_{l}~\tilde{Q}^{\nu}[\cdot]-\xi^{\bar{\nu}\ast}_{l}[\cdot]~\tilde{Q}^{\nu}.
\end{aligned}
\end{equation}\\
\section*{Acknowledgement}
This work is supported by the National Center for Theoretical Sciences and National Science and Technology Council, Taiwan, Grants No. NSTC 112-2123-M-006-001. N. L. is supported by the RIKEN Incentive Research Program and by MEXT KAKENHI Grant Numbers JP24H00816, JP24H00820. A.M. was supported by the Polish National Science Centre (NCN) under the Maestro Grant No. DEC-2019/34/A/ST2/00081. F.N. is supported in part by: Nippon Telegraph and Telephone Corporation (NTT) Research, the Japan Science and Technology Agency (JST) [via the Quantum Leap Flagship Program (Q-LEAP), and the Moonshot R\&D Grant Number JPMJMS2061],
the Asian Office of Aerospace Research and Development (AOARD) (via Grant No. FA2386-20-1-4069), and the Office of Naval Research (ONR) Global (via Grant No. N62909-23-1-2074).
\appendix

\bibliography{ref}

\begin{thebibliography}{75}%
\makeatletter
\providecommand \@ifxundefined [1]{%
 \@ifx{#1\undefined}
}%
\providecommand \@ifnum [1]{%
 \ifnum #1\expandafter \@firstoftwo
 \else \expandafter \@secondoftwo
 \fi
}%
\providecommand \@ifx [1]{%
 \ifx #1\expandafter \@firstoftwo
 \else \expandafter \@secondoftwo
 \fi
}%
\providecommand \natexlab [1]{#1}%
\providecommand \enquote  [1]{``#1''}%
\providecommand \bibnamefont  [1]{#1}%
\providecommand \bibfnamefont [1]{#1}%
\providecommand \citenamefont [1]{#1}%
\providecommand \href@noop [0]{\@secondoftwo}%
\providecommand \href [0]{\begingroup \@sanitize@url \@href}%
\providecommand \@href[1]{\@@startlink{#1}\@@href}%
\providecommand \@@href[1]{\endgroup#1\@@endlink}%
\providecommand \@sanitize@url [0]{\catcode `\\12\catcode `\$12\catcode `\&12\catcode `\#12\catcode `\^12\catcode `\_12\catcode `\%12\relax}%
\providecommand \@@startlink[1]{}%
\providecommand \@@endlink[0]{}%
\providecommand \url  [0]{\begingroup\@sanitize@url \@url }%
\providecommand \@url [1]{\endgroup\@href {#1}{\urlprefix }}%
\providecommand \urlprefix  [0]{URL }%
\providecommand \Eprint [0]{\href }%
\providecommand \doibase [0]{https://doi.org/}%
\providecommand \selectlanguage [0]{\@gobble}%
\providecommand \bibinfo  [0]{\@secondoftwo}%
\providecommand \bibfield  [0]{\@secondoftwo}%
\providecommand \translation [1]{[#1]}%
\providecommand \BibitemOpen [0]{}%
\providecommand \bibitemStop [0]{}%
\providecommand \bibitemNoStop [0]{.\EOS\space}%
\providecommand \EOS [0]{\spacefactor3000\relax}%
\providecommand \BibitemShut  [1]{\csname bibitem#1\endcsname}%
\let\auto@bib@innerbib\@empty
\bibitem [{\citenamefont {Heiss}(2012)}]{heiss2012physics}%
  \BibitemOpen
  \bibfield  {author} {\bibinfo {author} {\bibfnamefont {W.}~\bibnamefont {Heiss}},\ }\bibfield  {title} {\bibinfo {title} {The physics of exceptional points},\ }\href {https://iopscience.iop.org/article/10.1088/1751-8113/45/44/444016} {\bibfield  {journal} {\bibinfo  {journal} {J. Phys. A Math. Theor.}\ }\textbf {\bibinfo {volume} {45}},\ \bibinfo {pages} {444016} (\bibinfo {year} {2012})}\BibitemShut {NoStop}%
\bibitem [{\citenamefont {El-Ganainy}\ \emph {et~al.}(2018)\citenamefont {El-Ganainy}, \citenamefont {Makris}, \citenamefont {Khajavikhan}, \citenamefont {Musslimani}, \citenamefont {Rotter},\ and\ \citenamefont {Christodoulides}}]{el2018non}%
  \BibitemOpen
  \bibfield  {author} {\bibinfo {author} {\bibfnamefont {R.}~\bibnamefont {El-Ganainy}}, \bibinfo {author} {\bibfnamefont {K.~G.}\ \bibnamefont {Makris}}, \bibinfo {author} {\bibfnamefont {M.}~\bibnamefont {Khajavikhan}}, \bibinfo {author} {\bibfnamefont {Z.~H.}\ \bibnamefont {Musslimani}}, \bibinfo {author} {\bibfnamefont {S.}~\bibnamefont {Rotter}},\ and\ \bibinfo {author} {\bibfnamefont {D.~N.}\ \bibnamefont {Christodoulides}},\ }\bibfield  {title} {\bibinfo {title} {Non-{H}ermitian physics and $\mathcal{PT}$ symmetry},\ }\href {https://www.nature.com/articles/nphys4323} {\bibfield  {journal} {\bibinfo  {journal} {Nat. Phys.}\ }\textbf {\bibinfo {volume} {14}},\ \bibinfo {pages} {11} (\bibinfo {year} {2018})}\BibitemShut {NoStop}%
\bibitem [{\citenamefont {Miri}\ and\ \citenamefont {Alu}(2019)}]{miri2019exceptional}%
  \BibitemOpen
  \bibfield  {author} {\bibinfo {author} {\bibfnamefont {M.-A.}\ \bibnamefont {Miri}}\ and\ \bibinfo {author} {\bibfnamefont {A.}~\bibnamefont {Alu}},\ }\bibfield  {title} {\bibinfo {title} {Exceptional points in optics and photonics},\ }\href {https://www.science.org/doi/10.1126/science.aar7709} {\bibfield  {journal} {\bibinfo  {journal} {Science}\ }\textbf {\bibinfo {volume} {363}},\ \bibinfo {pages} {7709} (\bibinfo {year} {2019})}\BibitemShut {NoStop}%
\bibitem [{\citenamefont {{\"O}zdemir}\ \emph {et~al.}(2019)\citenamefont {{\"O}zdemir}, \citenamefont {Rotter} \emph {et~al.}}]{ozdemir2019parity}%
  \BibitemOpen
  \bibfield  {author} {\bibinfo {author} {\bibfnamefont {{\c{S}}.~K.}\ \bibnamefont {{\"O}zdemir}}, \bibinfo {author} {\bibfnamefont {S.}~\bibnamefont {Rotter}}, \emph {et~al.},\ }\bibfield  {title} {\bibinfo {title} {Parity--time symmetry and exceptional points in photonics},\ }\href {https://www.nature.com/articles/s41563-019-0304-9} {\bibfield  {journal} {\bibinfo  {journal} {Nat. Mater.}\ }\textbf {\bibinfo {volume} {18}},\ \bibinfo {pages} {783} (\bibinfo {year} {2019})}\BibitemShut {NoStop}%
\bibitem [{\citenamefont {Bender}\ and\ \citenamefont {Boettcher}(1998)}]{PhysRevLett.80.5243}%
  \BibitemOpen
  \bibfield  {author} {\bibinfo {author} {\bibfnamefont {C.~M.}\ \bibnamefont {Bender}}\ and\ \bibinfo {author} {\bibfnamefont {S.}~\bibnamefont {Boettcher}},\ }\bibfield  {title} {\bibinfo {title} {Real {S}pectra in {N}on-{H}ermitian {H}amiltonians {H}aving $\mathcal{P}\mathcal{T}$ {S}ymmetry},\ }\href {https://doi.org/10.1103/PhysRevLett.80.5243} {\bibfield  {journal} {\bibinfo  {journal} {Phys. Rev. Lett.}\ }\textbf {\bibinfo {volume} {80}},\ \bibinfo {pages} {5243} (\bibinfo {year} {1998})}\BibitemShut {NoStop}%
\bibitem [{\citenamefont {Bender}(2007)}]{bender2007making}%
  \BibitemOpen
  \bibfield  {author} {\bibinfo {author} {\bibfnamefont {C.~M.}\ \bibnamefont {Bender}},\ }\bibfield  {title} {\bibinfo {title} {Making sense of non-{H}ermitian {H}amiltonians},\ }\href {https://iopscience.iop.org/article/10.1088/0034-4885/70/6/R03/meta} {\bibfield  {journal} {\bibinfo  {journal} {Rep. Prog. Phys.}\ }\textbf {\bibinfo {volume} {70}},\ \bibinfo {pages} {947} (\bibinfo {year} {2007})}\BibitemShut {NoStop}%
\bibitem [{\citenamefont {Moiseyev}(2011)}]{moiseyev2011non}%
  \BibitemOpen
  \bibfield  {author} {\bibinfo {author} {\bibfnamefont {N.}~\bibnamefont {Moiseyev}},\ }\href@noop {} {\emph {\bibinfo {title} {Non-Hermitian quantum mechanics}}}\ (\bibinfo  {publisher} {Cambridge University Press},\ \bibinfo {year} {2011})\BibitemShut {NoStop}%
\bibitem [{\citenamefont {Mostafazadeh}(2002)}]{mostafazadeh2002pseudo}%
  \BibitemOpen
  \bibfield  {author} {\bibinfo {author} {\bibfnamefont {A.}~\bibnamefont {Mostafazadeh}},\ }\bibfield  {title} {\bibinfo {title} {Pseudo-{H}ermiticity versus {PT} symmetry: the necessary condition for the reality of the spectrum of a non-{H}ermitian {H}amiltonian},\ }\href {https://pubs.aip.org/aip/jmp/article/43/1/205/231882/Pseudo-Hermiticity-versus-PT-symmetry-The} {\bibfield  {journal} {\bibinfo  {journal} {J. Math. Phys.}\ }\textbf {\bibinfo {volume} {43}},\ \bibinfo {pages} {205} (\bibinfo {year} {2002})}\BibitemShut {NoStop}%
\bibitem [{\citenamefont {Peng}\ \emph {et~al.}(2014)\citenamefont {Peng}, \citenamefont {{\"O}zdemir}, \citenamefont {Rotter} \emph {et~al.}}]{peng2014loss}%
  \BibitemOpen
  \bibfield  {author} {\bibinfo {author} {\bibfnamefont {B.}~\bibnamefont {Peng}}, \bibinfo {author} {\bibfnamefont {{\c{S}}.}~\bibnamefont {{\"O}zdemir}}, \bibinfo {author} {\bibfnamefont {S.}~\bibnamefont {Rotter}}, \emph {et~al.},\ }\bibfield  {title} {\bibinfo {title} {Loss-induced suppression and revival of lasing},\ }\href {https://www.science.org/doi/10.1126/science.1258004} {\bibfield  {journal} {\bibinfo  {journal} {Science}\ }\textbf {\bibinfo {volume} {346}},\ \bibinfo {pages} {328} (\bibinfo {year} {2014})}\BibitemShut {NoStop}%
\bibitem [{\citenamefont {Jing}\ \emph {et~al.}(2014)\citenamefont {Jing} \emph {et~al.}}]{PhysRevLett.113.053604}%
  \BibitemOpen
  \bibfield  {author} {\bibinfo {author} {\bibfnamefont {H.}~\bibnamefont {Jing}} \emph {et~al.},\ }\bibfield  {title} {\bibinfo {title} {$\mathcal{PT}$-{S}ymmetric {P}honon {L}aser},\ }\href {https://doi.org/10.1103/PhysRevLett.113.053604} {\bibfield  {journal} {\bibinfo  {journal} {Phys. Rev. Lett.}\ }\textbf {\bibinfo {volume} {113}},\ \bibinfo {pages} {053604} (\bibinfo {year} {2014})}\BibitemShut {NoStop}%
\bibitem [{\citenamefont {Zhang}\ \emph {et~al.}(2018)\citenamefont {Zhang} \emph {et~al.}}]{zhang2018phonon}%
  \BibitemOpen
  \bibfield  {author} {\bibinfo {author} {\bibfnamefont {J.}~\bibnamefont {Zhang}} \emph {et~al.},\ }\bibfield  {title} {\bibinfo {title} {A phonon laser operating at an exceptional point},\ }\href {https://www.nature.com/articles/s41566-018-0213-5} {\bibfield  {journal} {\bibinfo  {journal} {Nat. Photon.}\ }\textbf {\bibinfo {volume} {12}},\ \bibinfo {pages} {479} (\bibinfo {year} {2018})}\BibitemShut {NoStop}%
\bibitem [{\citenamefont {Arkhipov}\ \emph {et~al.}(2023)\citenamefont {Arkhipov} \emph {et~al.}}]{arkhipov2023dynamically}%
  \BibitemOpen
  \bibfield  {author} {\bibinfo {author} {\bibfnamefont {I.~I.}\ \bibnamefont {Arkhipov}} \emph {et~al.},\ }\bibfield  {title} {\bibinfo {title} {Dynamically crossing diabolic points while encircling exceptional curves: {A} programmable symmetric-asymmetric multimode switch},\ }\href {https://www.nature.com/articles/s41467-023-37275-5} {\bibfield  {journal} {\bibinfo  {journal} {Nat. Commun.}\ }\textbf {\bibinfo {volume} {14}},\ \bibinfo {pages} {2076} (\bibinfo {year} {2023})}\BibitemShut {NoStop}%
\bibitem [{\citenamefont {Chen}\ \emph {et~al.}(2017)\citenamefont {Chen}, \citenamefont {Kaya~{\"O}zdemir}, \citenamefont {Zhao}, \citenamefont {Wiersig},\ and\ \citenamefont {Yang}}]{chen2017exceptional}%
  \BibitemOpen
  \bibfield  {author} {\bibinfo {author} {\bibfnamefont {W.}~\bibnamefont {Chen}}, \bibinfo {author} {\bibfnamefont {{\c{S}}.}~\bibnamefont {Kaya~{\"O}zdemir}}, \bibinfo {author} {\bibfnamefont {G.}~\bibnamefont {Zhao}}, \bibinfo {author} {\bibfnamefont {J.}~\bibnamefont {Wiersig}},\ and\ \bibinfo {author} {\bibfnamefont {L.}~\bibnamefont {Yang}},\ }\bibfield  {title} {\bibinfo {title} {Exceptional points enhance sensing in an optical microcavity},\ }\href {https://www.nature.com/articles/nature23281} {\bibfield  {journal} {\bibinfo  {journal} {Nature}\ }\textbf {\bibinfo {volume} {548}},\ \bibinfo {pages} {192} (\bibinfo {year} {2017})}\BibitemShut {NoStop}%
\bibitem [{\citenamefont {Hodaei}\ \emph {et~al.}(2017)\citenamefont {Hodaei}, \citenamefont {Hassan}, \citenamefont {Wittek}, \citenamefont {Garcia-Gracia}, \citenamefont {El-Ganainy}, \citenamefont {Christodoulides},\ and\ \citenamefont {Khajavikhan}}]{hodaei2017enhanced}%
  \BibitemOpen
  \bibfield  {author} {\bibinfo {author} {\bibfnamefont {H.}~\bibnamefont {Hodaei}}, \bibinfo {author} {\bibfnamefont {A.~U.}\ \bibnamefont {Hassan}}, \bibinfo {author} {\bibfnamefont {S.}~\bibnamefont {Wittek}}, \bibinfo {author} {\bibfnamefont {H.}~\bibnamefont {Garcia-Gracia}}, \bibinfo {author} {\bibfnamefont {R.}~\bibnamefont {El-Ganainy}}, \bibinfo {author} {\bibfnamefont {D.~N.}\ \bibnamefont {Christodoulides}},\ and\ \bibinfo {author} {\bibfnamefont {M.}~\bibnamefont {Khajavikhan}},\ }\bibfield  {title} {\bibinfo {title} {Enhanced sensitivity at higher-order exceptional points},\ }\href {https://www.nature.com/articles/nature23280} {\bibfield  {journal} {\bibinfo  {journal} {Nature}\ }\textbf {\bibinfo {volume} {548}},\ \bibinfo {pages} {187} (\bibinfo {year} {2017})}\BibitemShut {NoStop}%
\bibitem [{\citenamefont {Rechtsman}(2017)}]{rechtsman2017optical}%
  \BibitemOpen
  \bibfield  {author} {\bibinfo {author} {\bibfnamefont {M.~C.}\ \bibnamefont {Rechtsman}},\ }\bibfield  {title} {\bibinfo {title} {Optical sensing gets exceptional},\ }\href {https://www.nature.com/articles/548161a} {\bibfield  {journal} {\bibinfo  {journal} {Nature}\ }\textbf {\bibinfo {volume} {548}},\ \bibinfo {pages} {161} (\bibinfo {year} {2017})}\BibitemShut {NoStop}%
\bibitem [{\citenamefont {Wiersig}(2020)}]{wiersig2020review}%
  \BibitemOpen
  \bibfield  {author} {\bibinfo {author} {\bibfnamefont {J.}~\bibnamefont {Wiersig}},\ }\bibfield  {title} {\bibinfo {title} {Review of exceptional point-based sensors},\ }\href {https://opg.optica.org/prj/fulltext.cfm?uri=prj-8-9-1457&id=434541} {\bibfield  {journal} {\bibinfo  {journal} {Photon. Res.}\ }\textbf {\bibinfo {volume} {8}},\ \bibinfo {pages} {1457} (\bibinfo {year} {2020})}\BibitemShut {NoStop}%
\bibitem [{\citenamefont {Kuo}\ \emph {et~al.}(2020)\citenamefont {Kuo}, \citenamefont {Lambert}, \citenamefont {Miranowicz}, \citenamefont {Chen}, \citenamefont {Chen}, \citenamefont {Chen},\ and\ \citenamefont {Nori}}]{PhysRevA.101.013814}%
  \BibitemOpen
  \bibfield  {author} {\bibinfo {author} {\bibfnamefont {P.-C.}\ \bibnamefont {Kuo}}, \bibinfo {author} {\bibfnamefont {N.}~\bibnamefont {Lambert}}, \bibinfo {author} {\bibfnamefont {A.}~\bibnamefont {Miranowicz}}, \bibinfo {author} {\bibfnamefont {H.-B.}\ \bibnamefont {Chen}}, \bibinfo {author} {\bibfnamefont {G.-Y.}\ \bibnamefont {Chen}}, \bibinfo {author} {\bibfnamefont {Y.-N.}\ \bibnamefont {Chen}},\ and\ \bibinfo {author} {\bibfnamefont {F.}~\bibnamefont {Nori}},\ }\bibfield  {title} {\bibinfo {title} {Collectively induced exceptional points of quantum emitters coupled to nanoparticle surface plasmons},\ }\href {https://doi.org/10.1103/PhysRevA.101.013814} {\bibfield  {journal} {\bibinfo  {journal} {Phys. Rev. A}\ }\textbf {\bibinfo {volume} {101}},\ \bibinfo {pages} {013814} (\bibinfo {year} {2020})}\BibitemShut {NoStop}%
\bibitem [{\citenamefont {Minganti}\ \emph {et~al.}(2019)\citenamefont {Minganti}, \citenamefont {Miranowicz}, \citenamefont {Chhajlany},\ and\ \citenamefont {Nori}}]{PhysRevA.100.062131}%
  \BibitemOpen
  \bibfield  {author} {\bibinfo {author} {\bibfnamefont {F.}~\bibnamefont {Minganti}}, \bibinfo {author} {\bibfnamefont {A.}~\bibnamefont {Miranowicz}}, \bibinfo {author} {\bibfnamefont {R.~W.}\ \bibnamefont {Chhajlany}},\ and\ \bibinfo {author} {\bibfnamefont {F.}~\bibnamefont {Nori}},\ }\bibfield  {title} {\bibinfo {title} {Quantum exceptional points of non-{H}ermitian {H}amiltonians and {L}iouvillians: {T}he effects of quantum jumps},\ }\href {https://doi.org/10.1103/PhysRevA.100.062131} {\bibfield  {journal} {\bibinfo  {journal} {Phys. Rev. A}\ }\textbf {\bibinfo {volume} {100}},\ \bibinfo {pages} {062131} (\bibinfo {year} {2019})}\BibitemShut {NoStop}%
\bibitem [{\citenamefont {Arkhipov}\ \emph {et~al.}(2020{\natexlab{a}})\citenamefont {Arkhipov}, \citenamefont {Miranowicz}, \citenamefont {Minganti},\ and\ \citenamefont {Nori}}]{PhysRevA.101.013812}%
  \BibitemOpen
  \bibfield  {author} {\bibinfo {author} {\bibfnamefont {I.~I.}\ \bibnamefont {Arkhipov}}, \bibinfo {author} {\bibfnamefont {A.}~\bibnamefont {Miranowicz}}, \bibinfo {author} {\bibfnamefont {F.}~\bibnamefont {Minganti}},\ and\ \bibinfo {author} {\bibfnamefont {F.}~\bibnamefont {Nori}},\ }\bibfield  {title} {\bibinfo {title} {Quantum and semiclassical exceptional points of a linear system of coupled cavities with losses and gain within the {S}cully-{L}amb laser theory},\ }\href {https://doi.org/10.1103/PhysRevA.101.013812} {\bibfield  {journal} {\bibinfo  {journal} {Phys. Rev. A}\ }\textbf {\bibinfo {volume} {101}},\ \bibinfo {pages} {013812} (\bibinfo {year} {2020}{\natexlab{a}})}\BibitemShut {NoStop}%
\bibitem [{\citenamefont {Minganti}\ \emph {et~al.}(2020)\citenamefont {Minganti}, \citenamefont {Miranowicz}, \citenamefont {Chhajlany}, \citenamefont {Arkhipov},\ and\ \citenamefont {Nori}}]{PhysRevA.101.062112}%
  \BibitemOpen
  \bibfield  {author} {\bibinfo {author} {\bibfnamefont {F.}~\bibnamefont {Minganti}}, \bibinfo {author} {\bibfnamefont {A.}~\bibnamefont {Miranowicz}}, \bibinfo {author} {\bibfnamefont {R.~W.}\ \bibnamefont {Chhajlany}}, \bibinfo {author} {\bibfnamefont {I.~I.}\ \bibnamefont {Arkhipov}},\ and\ \bibinfo {author} {\bibfnamefont {F.}~\bibnamefont {Nori}},\ }\bibfield  {title} {\bibinfo {title} {Hybrid-{L}iouvillian formalism connecting exceptional points of non-{H}ermitian {H}amiltonians and {L}iouvillians via postselection of quantum trajectories},\ }\href {https://doi.org/10.1103/PhysRevA.101.062112} {\bibfield  {journal} {\bibinfo  {journal} {Phys. Rev. A}\ }\textbf {\bibinfo {volume} {101}},\ \bibinfo {pages} {062112} (\bibinfo {year} {2020})}\BibitemShut {NoStop}%
\bibitem [{\citenamefont {Arkhipov}\ \emph {et~al.}(2020{\natexlab{b}})\citenamefont {Arkhipov}, \citenamefont {Miranowicz}, \citenamefont {Minganti},\ and\ \citenamefont {Nori}}]{PhysRevA.102.033715}%
  \BibitemOpen
  \bibfield  {author} {\bibinfo {author} {\bibfnamefont {I.~I.}\ \bibnamefont {Arkhipov}}, \bibinfo {author} {\bibfnamefont {A.}~\bibnamefont {Miranowicz}}, \bibinfo {author} {\bibfnamefont {F.}~\bibnamefont {Minganti}},\ and\ \bibinfo {author} {\bibfnamefont {F.}~\bibnamefont {Nori}},\ }\bibfield  {title} {\bibinfo {title} {Liouvillian exceptional points of any order in dissipative linear bosonic systems: {C}oherence functions and switching between $\mathcal{PT}$ and anti-$\mathcal{PT}$ symmetries},\ }\href {https://doi.org/10.1103/PhysRevA.102.033715} {\bibfield  {journal} {\bibinfo  {journal} {Phys. Rev. A}\ }\textbf {\bibinfo {volume} {102}},\ \bibinfo {pages} {033715} (\bibinfo {year} {2020}{\natexlab{b}})}\BibitemShut {NoStop}%
\bibitem [{\citenamefont {Arkhipov}\ \emph {et~al.}(2021)\citenamefont {Arkhipov}, \citenamefont {Minganti}, \citenamefont {Miranowicz},\ and\ \citenamefont {Nori}}]{PhysRevA.104.012205}%
  \BibitemOpen
  \bibfield  {author} {\bibinfo {author} {\bibfnamefont {I.~I.}\ \bibnamefont {Arkhipov}}, \bibinfo {author} {\bibfnamefont {F.}~\bibnamefont {Minganti}}, \bibinfo {author} {\bibfnamefont {A.}~\bibnamefont {Miranowicz}},\ and\ \bibinfo {author} {\bibfnamefont {F.}~\bibnamefont {Nori}},\ }\bibfield  {title} {\bibinfo {title} {Generating high-order quantum exceptional points in synthetic dimensions},\ }\href {https://doi.org/10.1103/PhysRevA.104.012205} {\bibfield  {journal} {\bibinfo  {journal} {Phys. Rev. A}\ }\textbf {\bibinfo {volume} {104}},\ \bibinfo {pages} {012205} (\bibinfo {year} {2021})}\BibitemShut {NoStop}%
\bibitem [{\citenamefont {Khandelwal}\ \emph {et~al.}(2021)\citenamefont {Khandelwal}, \citenamefont {Brunner},\ and\ \citenamefont {Haack}}]{PRXQuantum.2.040346}%
  \BibitemOpen
  \bibfield  {author} {\bibinfo {author} {\bibfnamefont {S.}~\bibnamefont {Khandelwal}}, \bibinfo {author} {\bibfnamefont {N.}~\bibnamefont {Brunner}},\ and\ \bibinfo {author} {\bibfnamefont {G.}~\bibnamefont {Haack}},\ }\bibfield  {title} {\bibinfo {title} {Signatures of {L}iouvillian exceptional points in a quantum thermal machine},\ }\href {https://doi.org/10.1103/PRXQuantum.2.040346} {\bibfield  {journal} {\bibinfo  {journal} {PRX Quantum}\ }\textbf {\bibinfo {volume} {2}},\ \bibinfo {pages} {040346} (\bibinfo {year} {2021})}\BibitemShut {NoStop}%
\bibitem [{\citenamefont {Zhang}\ \emph {et~al.}(2022)\citenamefont {Zhang}, \citenamefont {Zhang}, \citenamefont {Ding}, \citenamefont {Li}, \citenamefont {Bu}, \citenamefont {Wang}, \citenamefont {Yan}, \citenamefont {Su}, \citenamefont {Chen}, \citenamefont {Nori} \emph {et~al.}}]{zhang2022dynamical}%
  \BibitemOpen
  \bibfield  {author} {\bibinfo {author} {\bibfnamefont {J.-W.}\ \bibnamefont {Zhang}}, \bibinfo {author} {\bibfnamefont {J.-Q.}\ \bibnamefont {Zhang}}, \bibinfo {author} {\bibfnamefont {G.-Y.}\ \bibnamefont {Ding}}, \bibinfo {author} {\bibfnamefont {J.-C.}\ \bibnamefont {Li}}, \bibinfo {author} {\bibfnamefont {J.-T.}\ \bibnamefont {Bu}}, \bibinfo {author} {\bibfnamefont {B.}~\bibnamefont {Wang}}, \bibinfo {author} {\bibfnamefont {L.-L.}\ \bibnamefont {Yan}}, \bibinfo {author} {\bibfnamefont {S.-L.}\ \bibnamefont {Su}}, \bibinfo {author} {\bibfnamefont {L.}~\bibnamefont {Chen}}, \bibinfo {author} {\bibfnamefont {F.}~\bibnamefont {Nori}}, \emph {et~al.},\ }\bibfield  {title} {\bibinfo {title} {Dynamical control of quantum heat engines using exceptional points},\ }\href {https://www.nature.com/articles/s41467-022-33667-1} {\bibfield  {journal} {\bibinfo  {journal} {Nat. Commun}\ }\textbf {\bibinfo {volume} {13}},\ \bibinfo {pages} {6225} (\bibinfo {year} {2022})}\BibitemShut {NoStop}%
\bibitem [{\citenamefont {Perina~Jr}\ \emph {et~al.}(2022)\citenamefont {Perina~Jr}, \citenamefont {Miranowicz}, \citenamefont {Chimczak},\ and\ \citenamefont {Kowalewska-Kudlaszyk}}]{perina2022quantum}%
  \BibitemOpen
  \bibfield  {author} {\bibinfo {author} {\bibfnamefont {J.}~\bibnamefont {Perina~Jr}}, \bibinfo {author} {\bibfnamefont {A.}~\bibnamefont {Miranowicz}}, \bibinfo {author} {\bibfnamefont {G.}~\bibnamefont {Chimczak}},\ and\ \bibinfo {author} {\bibfnamefont {A.}~\bibnamefont {Kowalewska-Kudlaszyk}},\ }\bibfield  {title} {\bibinfo {title} {Quantum {L}iouvillian exceptional and diabolical points for bosonic fields with quadratic {H}amiltonians: The {H}eisenberg-{L}angevin equation approach},\ }\href {https://quantum-journal.org/papers/q-2022-12-22-883/} {\bibfield  {journal} {\bibinfo  {journal} {Quantum}\ }\textbf {\bibinfo {volume} {6}},\ \bibinfo {pages} {883} (\bibinfo {year} {2022})}\BibitemShut {NoStop}%
\bibitem [{\citenamefont {Han}\ \emph {et~al.}(2023)\citenamefont {Han}, \citenamefont {Wu}, \citenamefont {Huang}, \citenamefont {Wu}, \citenamefont {Zou}, \citenamefont {Yi}, \citenamefont {Zhang}, \citenamefont {Li}, \citenamefont {Xu}, \citenamefont {Zheng}, \citenamefont {Fan}, \citenamefont {Wen}, \citenamefont {Yang},\ and\ \citenamefont {Zheng}}]{PhysRevLett.131.260201}%
  \BibitemOpen
  \bibfield  {author} {\bibinfo {author} {\bibfnamefont {P.-R.}\ \bibnamefont {Han}}, \bibinfo {author} {\bibfnamefont {F.}~\bibnamefont {Wu}}, \bibinfo {author} {\bibfnamefont {X.-J.}\ \bibnamefont {Huang}}, \bibinfo {author} {\bibfnamefont {H.-Z.}\ \bibnamefont {Wu}}, \bibinfo {author} {\bibfnamefont {C.-L.}\ \bibnamefont {Zou}}, \bibinfo {author} {\bibfnamefont {W.}~\bibnamefont {Yi}}, \bibinfo {author} {\bibfnamefont {M.}~\bibnamefont {Zhang}}, \bibinfo {author} {\bibfnamefont {H.}~\bibnamefont {Li}}, \bibinfo {author} {\bibfnamefont {K.}~\bibnamefont {Xu}}, \bibinfo {author} {\bibfnamefont {D.}~\bibnamefont {Zheng}}, \bibinfo {author} {\bibfnamefont {H.}~\bibnamefont {Fan}}, \bibinfo {author} {\bibfnamefont {J.}~\bibnamefont {Wen}}, \bibinfo {author} {\bibfnamefont {Z.-B.}\ \bibnamefont {Yang}},\ and\ \bibinfo {author} {\bibfnamefont {S.-B.}\ \bibnamefont {Zheng}},\ }\bibfield  {title} {\bibinfo {title} {Exceptional {E}ntanglement {P}henomena: {N}on-{H}ermiticity {M}eeting {N}onclassicality},\ }\href
  {https://doi.org/10.1103/PhysRevLett.131.260201} {\bibfield  {journal} {\bibinfo  {journal} {Phys. Rev. Lett.}\ }\textbf {\bibinfo {volume} {131}},\ \bibinfo {pages} {260201} (\bibinfo {year} {2023})}\BibitemShut {NoStop}%
\bibitem [{\citenamefont {Zhou}\ \emph {et~al.}(2023)\citenamefont {Zhou}, \citenamefont {Yu}, \citenamefont {Wu}, \citenamefont {Li}, \citenamefont {Zhang}, \citenamefont {Li},\ and\ \citenamefont {Chen}}]{PhysRevResearch.5.043036}%
  \BibitemOpen
  \bibfield  {author} {\bibinfo {author} {\bibfnamefont {Y.-L.}\ \bibnamefont {Zhou}}, \bibinfo {author} {\bibfnamefont {X.-D.}\ \bibnamefont {Yu}}, \bibinfo {author} {\bibfnamefont {C.-W.}\ \bibnamefont {Wu}}, \bibinfo {author} {\bibfnamefont {X.-Q.}\ \bibnamefont {Li}}, \bibinfo {author} {\bibfnamefont {J.}~\bibnamefont {Zhang}}, \bibinfo {author} {\bibfnamefont {W.}~\bibnamefont {Li}},\ and\ \bibinfo {author} {\bibfnamefont {P.-X.}\ \bibnamefont {Chen}},\ }\bibfield  {title} {\bibinfo {title} {Accelerating relaxation through {L}iouvillian exceptional point},\ }\href {https://doi.org/10.1103/PhysRevResearch.5.043036} {\bibfield  {journal} {\bibinfo  {journal} {Phys. Rev. Res.}\ }\textbf {\bibinfo {volume} {5}},\ \bibinfo {pages} {043036} (\bibinfo {year} {2023})}\BibitemShut {NoStop}%
\bibitem [{\citenamefont {Bu}\ \emph {et~al.}(2023)\citenamefont {Bu}, \citenamefont {Zhang}, \citenamefont {Ding} \emph {et~al.}}]{PhysRevLett.130.110402}%
  \BibitemOpen
  \bibfield  {author} {\bibinfo {author} {\bibfnamefont {J.-T.}\ \bibnamefont {Bu}}, \bibinfo {author} {\bibfnamefont {J.-Q.}\ \bibnamefont {Zhang}}, \bibinfo {author} {\bibfnamefont {G.-Y.}\ \bibnamefont {Ding}}, \emph {et~al.},\ }\bibfield  {title} {\bibinfo {title} {Enhancement of {Q}uantum {H}eat {E}ngine by {E}ncircling a {L}iouvillian {E}xceptional {P}oint},\ }\href {https://doi.org/10.1103/PhysRevLett.130.110402} {\bibfield  {journal} {\bibinfo  {journal} {Phys. Rev. Lett.}\ }\textbf {\bibinfo {volume} {130}},\ \bibinfo {pages} {110402} (\bibinfo {year} {2023})}\BibitemShut {NoStop}%
\bibitem [{\citenamefont {Khandelwal}\ \emph {et~al.}(2023)\citenamefont {Khandelwal}, \citenamefont {Chen}, \citenamefont {Murch},\ and\ \citenamefont {Haack}}]{khandelwal2023chiral}%
  \BibitemOpen
  \bibfield  {author} {\bibinfo {author} {\bibfnamefont {S.}~\bibnamefont {Khandelwal}}, \bibinfo {author} {\bibfnamefont {W.}~\bibnamefont {Chen}}, \bibinfo {author} {\bibfnamefont {K.~W.}\ \bibnamefont {Murch}},\ and\ \bibinfo {author} {\bibfnamefont {G.}~\bibnamefont {Haack}},\ }\bibfield  {title} {\bibinfo {title} {Chiral {B}ell-state transfer via dissipative {L}iouvillian dynamics},\ }\href {https://arxiv.org/abs/2310.11381} {\bibfield  {journal} {\bibinfo  {journal} {arXiv:2310.11381}\ } (\bibinfo {year} {2023})}\BibitemShut {NoStop}%
\bibitem [{\citenamefont {Abo}\ \emph {et~al.}(2024)\citenamefont {Abo}, \citenamefont {Tulewicz}, \citenamefont {Bartkiewicz}, \citenamefont {{\"O}zdemir},\ and\ \citenamefont {Miranowicz}}]{abo2024liouvillian}%
  \BibitemOpen
  \bibfield  {author} {\bibinfo {author} {\bibfnamefont {S.}~\bibnamefont {Abo}}, \bibinfo {author} {\bibfnamefont {P.}~\bibnamefont {Tulewicz}}, \bibinfo {author} {\bibfnamefont {K.}~\bibnamefont {Bartkiewicz}}, \bibinfo {author} {\bibfnamefont {{\c{S}}.~K.}\ \bibnamefont {{\"O}zdemir}},\ and\ \bibinfo {author} {\bibfnamefont {A.}~\bibnamefont {Miranowicz}},\ }\bibfield  {title} {\bibinfo {title} {Liouvillian {E}xceptional {P}oints of {N}on-{H}ermitian {S}ystems via {Q}uantum {P}rocess {T}omography},\ }\href {https://arxiv.org/abs/2401.14993} {\bibfield  {journal} {\bibinfo  {journal} {arXiv:2401.14993}\ } (\bibinfo {year} {2024})}\BibitemShut {NoStop}%
\bibitem [{\citenamefont {Laha}\ \emph {et~al.}(2024)\citenamefont {Laha}, \citenamefont {Miranowicz}, \citenamefont {Varshney},\ and\ \citenamefont {Ghosh}}]{PhysRevA.109.033511}%
  \BibitemOpen
  \bibfield  {author} {\bibinfo {author} {\bibfnamefont {A.}~\bibnamefont {Laha}}, \bibinfo {author} {\bibfnamefont {A.}~\bibnamefont {Miranowicz}}, \bibinfo {author} {\bibfnamefont {R.~K.}\ \bibnamefont {Varshney}},\ and\ \bibinfo {author} {\bibfnamefont {S.}~\bibnamefont {Ghosh}},\ }\bibfield  {title} {\bibinfo {title} {Correlated nonreciprocity around conjugate exceptional points},\ }\href {https://doi.org/10.1103/PhysRevA.109.033511} {\bibfield  {journal} {\bibinfo  {journal} {Phys. Rev. A}\ }\textbf {\bibinfo {volume} {109}},\ \bibinfo {pages} {033511} (\bibinfo {year} {2024})}\BibitemShut {NoStop}%
\bibitem [{\citenamefont {Breuer}\ and\ \citenamefont {Petruccione}(2002)}]{breuer2002theory}%
  \BibitemOpen
  \bibfield  {author} {\bibinfo {author} {\bibfnamefont {H.-P.}\ \bibnamefont {Breuer}}\ and\ \bibinfo {author} {\bibfnamefont {F.}~\bibnamefont {Petruccione}},\ }\href@noop {} {\emph {\bibinfo {title} {The theory of open quantum systems}}}\ (\bibinfo  {publisher} {Oxford University Press},\ \bibinfo {year} {2002})\BibitemShut {NoStop}%
\bibitem [{\citenamefont {Cheung}\ \emph {et~al.}(2018)\citenamefont {Cheung}, \citenamefont {Patil},\ and\ \citenamefont {Vengalattore}}]{PhysRevA.97.052116}%
  \BibitemOpen
  \bibfield  {author} {\bibinfo {author} {\bibfnamefont {H.~F.~H.}\ \bibnamefont {Cheung}}, \bibinfo {author} {\bibfnamefont {Y.~S.}\ \bibnamefont {Patil}},\ and\ \bibinfo {author} {\bibfnamefont {M.}~\bibnamefont {Vengalattore}},\ }\bibfield  {title} {\bibinfo {title} {Emergent phases and critical behavior in a non-{M}arkovian open quantum system},\ }\href {https://doi.org/10.1103/PhysRevA.97.052116} {\bibfield  {journal} {\bibinfo  {journal} {Phys. Rev. A}\ }\textbf {\bibinfo {volume} {97}},\ \bibinfo {pages} {052116} (\bibinfo {year} {2018})}\BibitemShut {NoStop}%
\bibitem [{\citenamefont {Garmon}\ \emph {et~al.}(2021)\citenamefont {Garmon}, \citenamefont {Ordonez},\ and\ \citenamefont {Hatano}}]{PhysRevResearch.3.033029}%
  \BibitemOpen
  \bibfield  {author} {\bibinfo {author} {\bibfnamefont {S.}~\bibnamefont {Garmon}}, \bibinfo {author} {\bibfnamefont {G.}~\bibnamefont {Ordonez}},\ and\ \bibinfo {author} {\bibfnamefont {N.}~\bibnamefont {Hatano}},\ }\bibfield  {title} {\bibinfo {title} {Anomalous-order exceptional point and non-{M}arkovian {P}urcell effect at threshold in one-dimensional continuum systems},\ }\href {https://doi.org/10.1103/PhysRevResearch.3.033029} {\bibfield  {journal} {\bibinfo  {journal} {Phys. Rev. Res.}\ }\textbf {\bibinfo {volume} {3}},\ \bibinfo {pages} {033029} (\bibinfo {year} {2021})}\BibitemShut {NoStop}%
\bibitem [{\citenamefont {Sergeev}\ \emph {et~al.}(2023)\citenamefont {Sergeev}, \citenamefont {Zyablovsky}, \citenamefont {Andrianov},\ and\ \citenamefont {Lozovik}}]{sergeev2023signature}%
  \BibitemOpen
  \bibfield  {author} {\bibinfo {author} {\bibfnamefont {T.}~\bibnamefont {Sergeev}}, \bibinfo {author} {\bibfnamefont {A.}~\bibnamefont {Zyablovsky}}, \bibinfo {author} {\bibfnamefont {E.}~\bibnamefont {Andrianov}},\ and\ \bibinfo {author} {\bibfnamefont {Y.~E.}\ \bibnamefont {Lozovik}},\ }\bibfield  {title} {\bibinfo {title} {Signature of exceptional point phase transition in {H}ermitian systems},\ }\href {https://quantum-journal.org/papers/q-2023-04-17-982/} {\bibfield  {journal} {\bibinfo  {journal} {Quantum}\ }\textbf {\bibinfo {volume} {7}},\ \bibinfo {pages} {982} (\bibinfo {year} {2023})}\BibitemShut {NoStop}%
\bibitem [{\citenamefont {Mouloudakis}\ and\ \citenamefont {Lambropoulos}(2022)}]{PhysRevA.106.053709}%
  \BibitemOpen
  \bibfield  {author} {\bibinfo {author} {\bibfnamefont {G.}~\bibnamefont {Mouloudakis}}\ and\ \bibinfo {author} {\bibfnamefont {P.}~\bibnamefont {Lambropoulos}},\ }\bibfield  {title} {\bibinfo {title} {Coalescence of non-{M}arkovian dissipation, quantum {Z}eno effect, and non-{H}ermitian physics in a simple realistic quantum system},\ }\href {https://doi.org/10.1103/PhysRevA.106.053709} {\bibfield  {journal} {\bibinfo  {journal} {Phys. Rev. A}\ }\textbf {\bibinfo {volume} {106}},\ \bibinfo {pages} {053709} (\bibinfo {year} {2022})}\BibitemShut {NoStop}%
\bibitem [{\citenamefont {Garraway}(1997)}]{PhysRevA.55.2290}%
  \BibitemOpen
  \bibfield  {author} {\bibinfo {author} {\bibfnamefont {B.~M.}\ \bibnamefont {Garraway}},\ }\bibfield  {title} {\bibinfo {title} {Nonperturbative decay of an atomic system in a cavity},\ }\href {https://doi.org/10.1103/PhysRevA.55.2290} {\bibfield  {journal} {\bibinfo  {journal} {Phys. Rev. A}\ }\textbf {\bibinfo {volume} {55}},\ \bibinfo {pages} {2290} (\bibinfo {year} {1997})}\BibitemShut {NoStop}%
\bibitem [{\citenamefont {Dalton}\ \emph {et~al.}(2001)\citenamefont {Dalton}, \citenamefont {Barnett},\ and\ \citenamefont {Garraway}}]{PhysRevA.64.053813}%
  \BibitemOpen
  \bibfield  {author} {\bibinfo {author} {\bibfnamefont {B.~J.}\ \bibnamefont {Dalton}}, \bibinfo {author} {\bibfnamefont {S.~M.}\ \bibnamefont {Barnett}},\ and\ \bibinfo {author} {\bibfnamefont {B.~M.}\ \bibnamefont {Garraway}},\ }\bibfield  {title} {\bibinfo {title} {Theory of pseudomodes in quantum optical processes},\ }\href {https://doi.org/10.1103/PhysRevA.64.053813} {\bibfield  {journal} {\bibinfo  {journal} {Phys. Rev. A}\ }\textbf {\bibinfo {volume} {64}},\ \bibinfo {pages} {053813} (\bibinfo {year} {2001})}\BibitemShut {NoStop}%
\bibitem [{\citenamefont {Mazzola}\ \emph {et~al.}(2009)\citenamefont {Mazzola}, \citenamefont {Maniscalco}, \citenamefont {Piilo}, \citenamefont {Suominen},\ and\ \citenamefont {Garraway}}]{PhysRevA.80.012104}%
  \BibitemOpen
  \bibfield  {author} {\bibinfo {author} {\bibfnamefont {L.}~\bibnamefont {Mazzola}}, \bibinfo {author} {\bibfnamefont {S.}~\bibnamefont {Maniscalco}}, \bibinfo {author} {\bibfnamefont {J.}~\bibnamefont {Piilo}}, \bibinfo {author} {\bibfnamefont {K.-A.}\ \bibnamefont {Suominen}},\ and\ \bibinfo {author} {\bibfnamefont {B.~M.}\ \bibnamefont {Garraway}},\ }\bibfield  {title} {\bibinfo {title} {Pseudomodes as an effective description of memory: {N}on-{M}arkovian dynamics of two-state systems in structured reservoirs},\ }\href {https://doi.org/10.1103/PhysRevA.80.012104} {\bibfield  {journal} {\bibinfo  {journal} {Phys. Rev. A}\ }\textbf {\bibinfo {volume} {80}},\ \bibinfo {pages} {012104} (\bibinfo {year} {2009})}\BibitemShut {NoStop}%
\bibitem [{\citenamefont {Pleasance}\ and\ \citenamefont {Petruccione}(2021)}]{pleasance2021pseudomode}%
  \BibitemOpen
  \bibfield  {author} {\bibinfo {author} {\bibfnamefont {G.}~\bibnamefont {Pleasance}}\ and\ \bibinfo {author} {\bibfnamefont {F.}~\bibnamefont {Petruccione}},\ }\bibfield  {title} {\bibinfo {title} {Pseudomode description of general open quantum system dynamics: non-perturbative master equation for the spin-boson model},\ }\href {https://arxiv.org/abs/2108.05755} {\bibfield  {journal} {\bibinfo  {journal} {arXiv:2108.05755}\ } (\bibinfo {year} {2021})}\BibitemShut {NoStop}%
\bibitem [{\citenamefont {Tamascelli}\ \emph {et~al.}(2018)\citenamefont {Tamascelli}, \citenamefont {Smirne}, \citenamefont {Huelga},\ and\ \citenamefont {Plenio}}]{PhysRevLett.120.030402}%
  \BibitemOpen
  \bibfield  {author} {\bibinfo {author} {\bibfnamefont {D.}~\bibnamefont {Tamascelli}}, \bibinfo {author} {\bibfnamefont {A.}~\bibnamefont {Smirne}}, \bibinfo {author} {\bibfnamefont {S.~F.}\ \bibnamefont {Huelga}},\ and\ \bibinfo {author} {\bibfnamefont {M.~B.}\ \bibnamefont {Plenio}},\ }\bibfield  {title} {\bibinfo {title} {Nonperturbative {T}reatment of non-{M}arkovian {D}ynamics of {O}pen {Q}uantum {S}ystems},\ }\href {https://doi.org/10.1103/PhysRevLett.120.030402} {\bibfield  {journal} {\bibinfo  {journal} {Phys. Rev. Lett.}\ }\textbf {\bibinfo {volume} {120}},\ \bibinfo {pages} {030402} (\bibinfo {year} {2018})}\BibitemShut {NoStop}%
\bibitem [{\citenamefont {Lambert}\ \emph {et~al.}(2019)\citenamefont {Lambert}, \citenamefont {Ahmed}, \citenamefont {Cirio},\ and\ \citenamefont {Nori}}]{Lambert2019}%
  \BibitemOpen
  \bibfield  {author} {\bibinfo {author} {\bibfnamefont {N.}~\bibnamefont {Lambert}}, \bibinfo {author} {\bibfnamefont {S.}~\bibnamefont {Ahmed}}, \bibinfo {author} {\bibfnamefont {M.}~\bibnamefont {Cirio}},\ and\ \bibinfo {author} {\bibfnamefont {F.}~\bibnamefont {Nori}},\ }\bibfield  {title} {\bibinfo {title} {Modelling the ultra-strongly coupled spin-boson model with unphysical modes},\ }\href {https://doi.org/10.1038/s41467-019-11656-1} {\bibfield  {journal} {\bibinfo  {journal} {Nat. Commun.}\ }\textbf {\bibinfo {volume} {10}},\ \bibinfo {pages} {3721} (\bibinfo {year} {2019})}\BibitemShut {NoStop}%
\bibitem [{\citenamefont {Pleasance}\ \emph {et~al.}(2020)\citenamefont {Pleasance}, \citenamefont {Garraway},\ and\ \citenamefont {Petruccione}}]{PhysRevResearch.2.043058}%
  \BibitemOpen
  \bibfield  {author} {\bibinfo {author} {\bibfnamefont {G.}~\bibnamefont {Pleasance}}, \bibinfo {author} {\bibfnamefont {B.~M.}\ \bibnamefont {Garraway}},\ and\ \bibinfo {author} {\bibfnamefont {F.}~\bibnamefont {Petruccione}},\ }\bibfield  {title} {\bibinfo {title} {Generalized theory of pseudomodes for exact descriptions of non-{M}arkovian quantum processes},\ }\href {https://doi.org/10.1103/PhysRevResearch.2.043058} {\bibfield  {journal} {\bibinfo  {journal} {Phys. Rev. Res.}\ }\textbf {\bibinfo {volume} {2}},\ \bibinfo {pages} {043058} (\bibinfo {year} {2020})}\BibitemShut {NoStop}%
\bibitem [{\citenamefont {Cirio}\ \emph {et~al.}(2023{\natexlab{a}})\citenamefont {Cirio}, \citenamefont {Lambert}, \citenamefont {Liang}, \citenamefont {Kuo}, \citenamefont {Chen}, \citenamefont {Menczel}, \citenamefont {Funo},\ and\ \citenamefont {Nori}}]{Mauro2023}%
  \BibitemOpen
  \bibfield  {author} {\bibinfo {author} {\bibfnamefont {M.}~\bibnamefont {Cirio}}, \bibinfo {author} {\bibfnamefont {N.}~\bibnamefont {Lambert}}, \bibinfo {author} {\bibfnamefont {P.}~\bibnamefont {Liang}}, \bibinfo {author} {\bibfnamefont {P.-C.}\ \bibnamefont {Kuo}}, \bibinfo {author} {\bibfnamefont {Y.-N.}\ \bibnamefont {Chen}}, \bibinfo {author} {\bibfnamefont {P.}~\bibnamefont {Menczel}}, \bibinfo {author} {\bibfnamefont {K.}~\bibnamefont {Funo}},\ and\ \bibinfo {author} {\bibfnamefont {F.}~\bibnamefont {Nori}},\ }\bibfield  {title} {\bibinfo {title} {Pseudofermion method for the exact description of fermionic environments: {F}rom single-molecule electronics to the {K}ondo resonance},\ }\href {https://doi.org/10.1103/PhysRevResearch.5.033011} {\bibfield  {journal} {\bibinfo  {journal} {Phys. Rev. Res.}\ }\textbf {\bibinfo {volume} {5}},\ \bibinfo {pages} {033011} (\bibinfo {year} {2023}{\natexlab{a}})}\BibitemShut {NoStop}%
\bibitem [{\citenamefont {Luo}\ \emph {et~al.}(2023)\citenamefont {Luo}, \citenamefont {Lambert}, \citenamefont {Liang},\ and\ \citenamefont {Cirio}}]{PRXQuantum.4.030316}%
  \BibitemOpen
  \bibfield  {author} {\bibinfo {author} {\bibfnamefont {S.}~\bibnamefont {Luo}}, \bibinfo {author} {\bibfnamefont {N.}~\bibnamefont {Lambert}}, \bibinfo {author} {\bibfnamefont {P.}~\bibnamefont {Liang}},\ and\ \bibinfo {author} {\bibfnamefont {M.}~\bibnamefont {Cirio}},\ }\bibfield  {title} {\bibinfo {title} {{Q}uantum-{C}lassical {D}ecomposition of {G}aussian {Q}uantum {E}nvironments: {A} {S}tochastic {P}seudomode {M}odel},\ }\href {https://doi.org/10.1103/PRXQuantum.4.030316} {\bibfield  {journal} {\bibinfo  {journal} {PRX Quantum}\ }\textbf {\bibinfo {volume} {4}},\ \bibinfo {pages} {030316} (\bibinfo {year} {2023})}\BibitemShut {NoStop}%
\bibitem [{\citenamefont {Cirio}\ \emph {et~al.}(2023{\natexlab{b}})\citenamefont {Cirio}, \citenamefont {Luo}, \citenamefont {Liang}, \citenamefont {Nori},\ and\ \citenamefont {Lambert}}]{cirio2023modeling}%
  \BibitemOpen
  \bibfield  {author} {\bibinfo {author} {\bibfnamefont {M.}~\bibnamefont {Cirio}}, \bibinfo {author} {\bibfnamefont {S.}~\bibnamefont {Luo}}, \bibinfo {author} {\bibfnamefont {P.}~\bibnamefont {Liang}}, \bibinfo {author} {\bibfnamefont {F.}~\bibnamefont {Nori}},\ and\ \bibinfo {author} {\bibfnamefont {N.}~\bibnamefont {Lambert}},\ }\bibfield  {title} {\bibinfo {title} {Modeling the unphysical pseudomode model with physical ensembles: simulation, mitigation, and restructuring of non-{M}arkovian quantum noise},\ }\href {https://arxiv.org/abs/2311.15240} {\bibfield  {journal} {\bibinfo  {journal} {arXiv:2311.15240}\ } (\bibinfo {year} {2023}{\natexlab{b}})}\BibitemShut {NoStop}%
\bibitem [{\citenamefont {Menczel}\ \emph {et~al.}(2024)\citenamefont {Menczel}, \citenamefont {Funo}, \citenamefont {Cirio}, \citenamefont {Lambert},\ and\ \citenamefont {Nori}}]{menczel2024non}%
  \BibitemOpen
  \bibfield  {author} {\bibinfo {author} {\bibfnamefont {P.}~\bibnamefont {Menczel}}, \bibinfo {author} {\bibfnamefont {K.}~\bibnamefont {Funo}}, \bibinfo {author} {\bibfnamefont {M.}~\bibnamefont {Cirio}}, \bibinfo {author} {\bibfnamefont {N.}~\bibnamefont {Lambert}},\ and\ \bibinfo {author} {\bibfnamefont {F.}~\bibnamefont {Nori}},\ }\bibfield  {title} {\bibinfo {title} {Non-{H}ermitian {P}seudomodes for {S}trongly {C}oupled {O}pen {Q}uantum {S}ystems: {U}nravelings, {C}orrelations and {T}hermodynamics},\ }\href {https://arxiv.org/abs/2401.11830} {\bibfield  {journal} {\bibinfo  {journal} {arXiv:2401.11830}\ } (\bibinfo {year} {2024})}\BibitemShut {NoStop}%
\bibitem [{\citenamefont {Jin}\ \emph {et~al.}(2008)\citenamefont {Jin}, \citenamefont {Zheng},\ and\ \citenamefont {Yan}}]{jin2008exact}%
  \BibitemOpen
  \bibfield  {author} {\bibinfo {author} {\bibfnamefont {J.}~\bibnamefont {Jin}}, \bibinfo {author} {\bibfnamefont {X.}~\bibnamefont {Zheng}},\ and\ \bibinfo {author} {\bibfnamefont {Y.}~\bibnamefont {Yan}},\ }\bibfield  {title} {\bibinfo {title} {Exact dynamics of dissipative electronic systems and quantum transport: Hierarchical equations of motion approach},\ }\href@noop {} {\bibfield  {journal} {\bibinfo  {journal} {J. Chem. Phys.}\ }\textbf {\bibinfo {volume} {128}} (\bibinfo {year} {2008})}\BibitemShut {NoStop}%
\bibitem [{\citenamefont {Kreisbeck}\ \emph {et~al.}(2011)\citenamefont {Kreisbeck}, \citenamefont {Kramer}, \citenamefont {Rodriguez},\ and\ \citenamefont {Hein}}]{kreisbeck2011high}%
  \BibitemOpen
  \bibfield  {author} {\bibinfo {author} {\bibfnamefont {C.}~\bibnamefont {Kreisbeck}}, \bibinfo {author} {\bibfnamefont {T.}~\bibnamefont {Kramer}}, \bibinfo {author} {\bibfnamefont {M.}~\bibnamefont {Rodriguez}},\ and\ \bibinfo {author} {\bibfnamefont {B.}~\bibnamefont {Hein}},\ }\bibfield  {title} {\bibinfo {title} {High-performance solution of hierarchical equations of motion for studying energy transfer in light-harvesting complexes},\ }\href {https://doi.org/10.1021/ct200126d} {\bibfield  {journal} {\bibinfo  {journal} {Journal of Chemical Theory and Computation}\ }\textbf {\bibinfo {volume} {7}},\ \bibinfo {pages} {2166} (\bibinfo {year} {2011})}\BibitemShut {NoStop}%
\bibitem [{\citenamefont {Ma}\ \emph {et~al.}(2012)\citenamefont {Ma}, \citenamefont {Sun}, \citenamefont {Wang},\ and\ \citenamefont {Nori}}]{PhysRevA.85.062323}%
  \BibitemOpen
  \bibfield  {author} {\bibinfo {author} {\bibfnamefont {J.}~\bibnamefont {Ma}}, \bibinfo {author} {\bibfnamefont {Z.}~\bibnamefont {Sun}}, \bibinfo {author} {\bibfnamefont {X.}~\bibnamefont {Wang}},\ and\ \bibinfo {author} {\bibfnamefont {F.}~\bibnamefont {Nori}},\ }\bibfield  {title} {\bibinfo {title} {Entanglement dynamics of two qubits in a common bath},\ }\href {https://doi.org/10.1103/PhysRevA.85.062323} {\bibfield  {journal} {\bibinfo  {journal} {Phys. Rev. A}\ }\textbf {\bibinfo {volume} {85}},\ \bibinfo {pages} {062323} (\bibinfo {year} {2012})}\BibitemShut {NoStop}%
\bibitem [{\citenamefont {Li}\ \emph {et~al.}(2012)\citenamefont {Li}, \citenamefont {Tong}, \citenamefont {Zheng}, \citenamefont {Hou}, \citenamefont {Wei}, \citenamefont {Hu},\ and\ \citenamefont {Yan}}]{Yan2012}%
  \BibitemOpen
  \bibfield  {author} {\bibinfo {author} {\bibfnamefont {Z.~H.}\ \bibnamefont {Li}}, \bibinfo {author} {\bibfnamefont {N.~H.}\ \bibnamefont {Tong}}, \bibinfo {author} {\bibfnamefont {X.}~\bibnamefont {Zheng}}, \bibinfo {author} {\bibfnamefont {D.}~\bibnamefont {Hou}}, \bibinfo {author} {\bibfnamefont {J.~H.}\ \bibnamefont {Wei}}, \bibinfo {author} {\bibfnamefont {J.}~\bibnamefont {Hu}},\ and\ \bibinfo {author} {\bibfnamefont {Y.~J.}\ \bibnamefont {Yan}},\ }\bibfield  {title} {\bibinfo {title} {Hierarchical {L}iouville-space approach for accurate and universal characterization of quantum impurity systems},\ }\href {https://doi.org/10.1103/PhysRevLett.109.266403} {\bibfield  {journal} {\bibinfo  {journal} {Phys. Rev. Lett.}\ }\textbf {\bibinfo {volume} {109}},\ \bibinfo {pages} {266403} (\bibinfo {year} {2012})}\BibitemShut {NoStop}%
\bibitem [{\citenamefont {Dunn}\ \emph {et~al.}(2019)\citenamefont {Dunn}, \citenamefont {Tempelaar},\ and\ \citenamefont {Reichman}}]{dunn2019removing}%
  \BibitemOpen
  \bibfield  {author} {\bibinfo {author} {\bibfnamefont {I.~S.}\ \bibnamefont {Dunn}}, \bibinfo {author} {\bibfnamefont {R.}~\bibnamefont {Tempelaar}},\ and\ \bibinfo {author} {\bibfnamefont {D.~R.}\ \bibnamefont {Reichman}},\ }\bibfield  {title} {\bibinfo {title} {Removing instabilities in the hierarchical equations of motion: {E}xact and approximate projection approaches},\ }\href {https://pubs.aip.org/aip/jcp/article/150/18/184109/198666/Removing-instabilities-in-the-hierarchical} {\bibfield  {journal} {\bibinfo  {journal} {J. Chem. Phys.}\ }\textbf {\bibinfo {volume} {150}} (\bibinfo {year} {2019})}\BibitemShut {NoStop}%
\bibitem [{\citenamefont {Tanimura}(2020)}]{Tanimura01}%
  \BibitemOpen
  \bibfield  {author} {\bibinfo {author} {\bibfnamefont {Y.}~\bibnamefont {Tanimura}},\ }\bibfield  {title} {\bibinfo {title} {Numerically {\textquotedblleft}exact{\textquotedblright} approach to open quantum dynamics: The hierarchical equations of motion ({HEOM})},\ }\href {https://doi.org/10.1063/5.0011599} {\bibfield  {journal} {\bibinfo  {journal} {J. Chem. Phys.}\ }\textbf {\bibinfo {volume} {153}},\ \bibinfo {pages} {020901} (\bibinfo {year} {2020})}\BibitemShut {NoStop}%
\bibitem [{\citenamefont {Ikeda}\ and\ \citenamefont {Scholes}(2020)}]{ikeda2020generalization}%
  \BibitemOpen
  \bibfield  {author} {\bibinfo {author} {\bibfnamefont {T.}~\bibnamefont {Ikeda}}\ and\ \bibinfo {author} {\bibfnamefont {G.~D.}\ \bibnamefont {Scholes}},\ }\bibfield  {title} {\bibinfo {title} {Generalization of the hierarchical equations of motion theory for efficient calculations with arbitrary correlation functions},\ }\href@noop {} {\bibfield  {journal} {\bibinfo  {journal} {J. Chem. Phys.}\ }\textbf {\bibinfo {volume} {152}} (\bibinfo {year} {2020})}\BibitemShut {NoStop}%
\bibitem [{\citenamefont {Cirio}\ \emph {et~al.}(2022)\citenamefont {Cirio}, \citenamefont {Kuo}, \citenamefont {Chen}, \citenamefont {Nori},\ and\ \citenamefont {Lambert}}]{Mauro2022}%
  \BibitemOpen
  \bibfield  {author} {\bibinfo {author} {\bibfnamefont {M.}~\bibnamefont {Cirio}}, \bibinfo {author} {\bibfnamefont {P.~C.}\ \bibnamefont {Kuo}}, \bibinfo {author} {\bibfnamefont {Y.~N.}\ \bibnamefont {Chen}}, \bibinfo {author} {\bibfnamefont {F.}~\bibnamefont {Nori}},\ and\ \bibinfo {author} {\bibfnamefont {N.}~\bibnamefont {Lambert}},\ }\bibfield  {title} {\bibinfo {title} {Canonical derivation of the fermionic influence superoperator},\ }\href {https://doi.org/10.1103/PhysRevB.105.035121} {\bibfield  {journal} {\bibinfo  {journal} {Phys. Rev. B}\ }\textbf {\bibinfo {volume} {105}},\ \bibinfo {pages} {035121} (\bibinfo {year} {2022})}\BibitemShut {NoStop}%
\bibitem [{\citenamefont {Lambert}\ \emph {et~al.}(2023)\citenamefont {Lambert}, \citenamefont {Raheja}, \citenamefont {Cross}, \citenamefont {Menczel}, \citenamefont {Ahmed}, \citenamefont {Pitchford}, \citenamefont {Burgarth},\ and\ \citenamefont {Nori}}]{lambert2020bofinheom}%
  \BibitemOpen
  \bibfield  {author} {\bibinfo {author} {\bibfnamefont {N.}~\bibnamefont {Lambert}}, \bibinfo {author} {\bibfnamefont {T.}~\bibnamefont {Raheja}}, \bibinfo {author} {\bibfnamefont {S.}~\bibnamefont {Cross}}, \bibinfo {author} {\bibfnamefont {P.}~\bibnamefont {Menczel}}, \bibinfo {author} {\bibfnamefont {S.}~\bibnamefont {Ahmed}}, \bibinfo {author} {\bibfnamefont {A.}~\bibnamefont {Pitchford}}, \bibinfo {author} {\bibfnamefont {D.}~\bibnamefont {Burgarth}},\ and\ \bibinfo {author} {\bibfnamefont {F.}~\bibnamefont {Nori}},\ }\bibfield  {title} {\bibinfo {title} {Qu{T}i{P}-{B}o{F}i{N}: A bosonic and fermionic numerical hierarchical-equations-of-motion library with applications in light-harvesting, quantum control, and single-molecule electronics},\ }\href {https://doi.org/10.1103/PhysRevResearch.5.013181} {\bibfield  {journal} {\bibinfo  {journal} {Phys. Rev. Res.}\ }\textbf {\bibinfo {volume} {5}},\ \bibinfo {pages} {013181} (\bibinfo {year} {2023})}\BibitemShut {NoStop}%
\bibitem [{\citenamefont {Huang}\ \emph {et~al.}(2023)\citenamefont {Huang}, \citenamefont {Kuo}, \citenamefont {Lambert}, \citenamefont {Cirio}, \citenamefont {Cross}, \citenamefont {Yang}, \citenamefont {Nori},\ and\ \citenamefont {Chen}}]{Huang2023}%
  \BibitemOpen
  \bibfield  {author} {\bibinfo {author} {\bibfnamefont {Y.-T.}\ \bibnamefont {Huang}}, \bibinfo {author} {\bibfnamefont {P.-C.}\ \bibnamefont {Kuo}}, \bibinfo {author} {\bibfnamefont {N.}~\bibnamefont {Lambert}}, \bibinfo {author} {\bibfnamefont {M.}~\bibnamefont {Cirio}}, \bibinfo {author} {\bibfnamefont {S.}~\bibnamefont {Cross}}, \bibinfo {author} {\bibfnamefont {S.-L.}\ \bibnamefont {Yang}}, \bibinfo {author} {\bibfnamefont {F.}~\bibnamefont {Nori}},\ and\ \bibinfo {author} {\bibfnamefont {Y.-N.}\ \bibnamefont {Chen}},\ }\bibfield  {title} {\bibinfo {title} {An efficient {J}ulia framework for hierarchical equations of motion in open quantum systems},\ }\href {http://dx.doi.org/10.1038/s42005-023-01427-2} {\bibfield  {journal} {\bibinfo  {journal} {Commun. Phys.}\ }\textbf {\bibinfo {volume} {6}},\ \bibinfo {pages} {313} (\bibinfo {year} {2023})}\BibitemShut {NoStop}%
\bibitem [{\citenamefont {Kuo}\ \emph {et~al.}(2023)\citenamefont {Kuo}, \citenamefont {Lambert}, \citenamefont {Cirio}, \citenamefont {Huang}, \citenamefont {Nori},\ and\ \citenamefont {Chen}}]{KondoQED2023}%
  \BibitemOpen
  \bibfield  {author} {\bibinfo {author} {\bibfnamefont {P.-C.}\ \bibnamefont {Kuo}}, \bibinfo {author} {\bibfnamefont {N.}~\bibnamefont {Lambert}}, \bibinfo {author} {\bibfnamefont {M.}~\bibnamefont {Cirio}}, \bibinfo {author} {\bibfnamefont {Y.-T.}\ \bibnamefont {Huang}}, \bibinfo {author} {\bibfnamefont {F.}~\bibnamefont {Nori}},\ and\ \bibinfo {author} {\bibfnamefont {Y.-N.}\ \bibnamefont {Chen}},\ }\bibfield  {title} {\bibinfo {title} {Kondo {QED}: The {K}ondo effect and photon trapping in a two-impurity {A}nderson model ultrastrongly coupled to light},\ }\href {https://doi.org/10.1103/PhysRevResearch.5.043177} {\bibfield  {journal} {\bibinfo  {journal} {Phys. Rev. Res.}\ }\textbf {\bibinfo {volume} {5}},\ \bibinfo {pages} {043177} (\bibinfo {year} {2023})}\BibitemShut {NoStop}%
\bibitem [{\citenamefont {Nakamura}\ and\ \citenamefont {Tanimura}(2021)}]{Tanimura02}%
  \BibitemOpen
  \bibfield  {author} {\bibinfo {author} {\bibfnamefont {K.}~\bibnamefont {Nakamura}}\ and\ \bibinfo {author} {\bibfnamefont {Y.}~\bibnamefont {Tanimura}},\ }\bibfield  {title} {\bibinfo {title} {{Optical response of laser-driven charge-transfer complex described by Holstein–Hubbard model coupled to heat baths: Hierarchical equations of motion approach}},\ }\href {https://doi.org/10.1063/5.0060208} {\bibfield  {journal} {\bibinfo  {journal} {J. Chem. Phys.}\ }\textbf {\bibinfo {volume} {155}},\ \bibinfo {pages} {064106} (\bibinfo {year} {2021})}\BibitemShut {NoStop}%
\bibitem [{\citenamefont {Kato}\ and\ \citenamefont {Tanimura}(2016)}]{Tanimura05}%
  \BibitemOpen
  \bibfield  {author} {\bibinfo {author} {\bibfnamefont {A.}~\bibnamefont {Kato}}\ and\ \bibinfo {author} {\bibfnamefont {Y.}~\bibnamefont {Tanimura}},\ }\bibfield  {title} {\bibinfo {title} {Quantum heat current under non-perturbative and non-{M}arkovian conditions: Applications to heat machines},\ }\href {https://doi.org/10.1063/1.4971370} {\bibfield  {journal} {\bibinfo  {journal} {J. Chem. Phys}\ }\textbf {\bibinfo {volume} {145}},\ \bibinfo {pages} {224105} (\bibinfo {year} {2016})}\BibitemShut {NoStop}%
\bibitem [{\citenamefont {Breuer}\ \emph {et~al.}(2009)\citenamefont {Breuer}, \citenamefont {Laine},\ and\ \citenamefont {Piilo}}]{PhysRevLett.103.210401}%
  \BibitemOpen
  \bibfield  {author} {\bibinfo {author} {\bibfnamefont {H.-P.}\ \bibnamefont {Breuer}}, \bibinfo {author} {\bibfnamefont {E.-M.}\ \bibnamefont {Laine}},\ and\ \bibinfo {author} {\bibfnamefont {J.}~\bibnamefont {Piilo}},\ }\bibfield  {title} {\bibinfo {title} {Measure for the {D}egree of {N}on-{M}arkovian {B}ehavior of {Q}uantum {P}rocesses in {O}pen {S}ystems},\ }\href {https://doi.org/10.1103/PhysRevLett.103.210401} {\bibfield  {journal} {\bibinfo  {journal} {Phys. Rev. Lett.}\ }\textbf {\bibinfo {volume} {103}},\ \bibinfo {pages} {210401} (\bibinfo {year} {2009})}\BibitemShut {NoStop}%
\bibitem [{\citenamefont {Rivas}\ \emph {et~al.}(2010)\citenamefont {Rivas}, \citenamefont {Huelga},\ and\ \citenamefont {Plenio}}]{PhysRevLett.105.050403}%
  \BibitemOpen
  \bibfield  {author} {\bibinfo {author} {\bibfnamefont {A.}~\bibnamefont {Rivas}}, \bibinfo {author} {\bibfnamefont {S.~F.}\ \bibnamefont {Huelga}},\ and\ \bibinfo {author} {\bibfnamefont {M.~B.}\ \bibnamefont {Plenio}},\ }\bibfield  {title} {\bibinfo {title} {Entanglement and {N}on-{M}arkovianity of {Q}uantum {E}volutions},\ }\href {https://doi.org/10.1103/PhysRevLett.105.050403} {\bibfield  {journal} {\bibinfo  {journal} {Phys. Rev. Lett.}\ }\textbf {\bibinfo {volume} {105}},\ \bibinfo {pages} {050403} (\bibinfo {year} {2010})}\BibitemShut {NoStop}%
\bibitem [{\citenamefont {Yan}(2014)}]{yan2014theory}%
  \BibitemOpen
  \bibfield  {author} {\bibinfo {author} {\bibfnamefont {Y.}~\bibnamefont {Yan}},\ }\bibfield  {title} {\bibinfo {title} {Theory of open quantum systems with bath of electrons and phonons and spins: {M}any-dissipaton density matrixes approach},\ }\href {https://pubs.aip.org/aip/jcp/article/140/5/054105/73714} {\bibfield  {journal} {\bibinfo  {journal} {J. Chem. Phys.}\ }\textbf {\bibinfo {volume} {140}} (\bibinfo {year} {2014})}\BibitemShut {NoStop}%
\bibitem [{\citenamefont {Wang}\ and\ \citenamefont {Yan}(2022)}]{wang2022quantum}%
  \BibitemOpen
  \bibfield  {author} {\bibinfo {author} {\bibfnamefont {Y.}~\bibnamefont {Wang}}\ and\ \bibinfo {author} {\bibfnamefont {Y.}~\bibnamefont {Yan}},\ }\bibfield  {title} {\bibinfo {title} {Quantum mechanics of open systems: {D}issipaton theories},\ }\href {https://pubs.aip.org/aip/jcp/article/157/17/170901/2841974} {\bibfield  {journal} {\bibinfo  {journal} {J. Chem. Phys.}\ }\textbf {\bibinfo {volume} {157}} (\bibinfo {year} {2022})}\BibitemShut {NoStop}%
\bibitem [{\citenamefont {Iles-Smith}\ \emph {et~al.}(2014)\citenamefont {Iles-Smith}, \citenamefont {Lambert},\ and\ \citenamefont {Nazir}}]{PhysRevA.90.032114}%
  \BibitemOpen
  \bibfield  {author} {\bibinfo {author} {\bibfnamefont {J.}~\bibnamefont {Iles-Smith}}, \bibinfo {author} {\bibfnamefont {N.}~\bibnamefont {Lambert}},\ and\ \bibinfo {author} {\bibfnamefont {A.}~\bibnamefont {Nazir}},\ }\bibfield  {title} {\bibinfo {title} {Environmental dynamics, correlations, and the emergence of noncanonical equilibrium states in open quantum systems},\ }\href {https://doi.org/10.1103/PhysRevA.90.032114} {\bibfield  {journal} {\bibinfo  {journal} {Phys. Rev. A}\ }\textbf {\bibinfo {volume} {90}},\ \bibinfo {pages} {032114} (\bibinfo {year} {2014})}\BibitemShut {NoStop}%
\bibitem [{\citenamefont {Iles-Smith}\ \emph {et~al.}(2016)\citenamefont {Iles-Smith}, \citenamefont {Dijkstra}, \citenamefont {Lambert},\ and\ \citenamefont {Nazir}}]{iles2016energy}%
  \BibitemOpen
  \bibfield  {author} {\bibinfo {author} {\bibfnamefont {J.}~\bibnamefont {Iles-Smith}}, \bibinfo {author} {\bibfnamefont {A.~G.}\ \bibnamefont {Dijkstra}}, \bibinfo {author} {\bibfnamefont {N.}~\bibnamefont {Lambert}},\ and\ \bibinfo {author} {\bibfnamefont {A.}~\bibnamefont {Nazir}},\ }\bibfield  {title} {\bibinfo {title} {Energy transfer in structured and unstructured environments: {M}aster equations beyond the {B}orn-{M}arkov approximations},\ }\href {https://iopscience.iop.org/article/10.1088/1367-2630/18/7/073007/meta} {\bibfield  {journal} {\bibinfo  {journal} {J. Chem. Phys.}\ }\textbf {\bibinfo {volume} {144}} (\bibinfo {year} {2016})}\BibitemShut {NoStop}%
\bibitem [{\citenamefont {Restrepo}\ \emph {et~al.}(2019)\citenamefont {Restrepo}, \citenamefont {B\"ohling}, \citenamefont {Cerrillo},\ and\ \citenamefont {Schaller}}]{PhysRevB.100.035109}%
  \BibitemOpen
  \bibfield  {author} {\bibinfo {author} {\bibfnamefont {S.}~\bibnamefont {Restrepo}}, \bibinfo {author} {\bibfnamefont {S.}~\bibnamefont {B\"ohling}}, \bibinfo {author} {\bibfnamefont {J.}~\bibnamefont {Cerrillo}},\ and\ \bibinfo {author} {\bibfnamefont {G.}~\bibnamefont {Schaller}},\ }\bibfield  {title} {\bibinfo {title} {Electron pumping in the strong coupling and non-{M}arkovian regime: {A} reaction coordinate mapping approach},\ }\href {https://doi.org/10.1103/PhysRevB.100.035109} {\bibfield  {journal} {\bibinfo  {journal} {Phys. Rev. B}\ }\textbf {\bibinfo {volume} {100}},\ \bibinfo {pages} {035109} (\bibinfo {year} {2019})}\BibitemShut {NoStop}%
\bibitem [{\citenamefont {Anto-Sztrikacs}\ and\ \citenamefont {Segal}(2021)}]{PhysRevA.104.052617}%
  \BibitemOpen
  \bibfield  {author} {\bibinfo {author} {\bibfnamefont {N.}~\bibnamefont {Anto-Sztrikacs}}\ and\ \bibinfo {author} {\bibfnamefont {D.}~\bibnamefont {Segal}},\ }\bibfield  {title} {\bibinfo {title} {Capturing non-{M}arkovian dynamics with the reaction coordinate method},\ }\href {https://doi.org/10.1103/PhysRevA.104.052617} {\bibfield  {journal} {\bibinfo  {journal} {Phys. Rev. A}\ }\textbf {\bibinfo {volume} {104}},\ \bibinfo {pages} {052617} (\bibinfo {year} {2021})}\BibitemShut {NoStop}%
\bibitem [{\citenamefont {Ding}\ \emph {et~al.}(2022)\citenamefont {Ding}, \citenamefont {Fang},\ and\ \citenamefont {Ma}}]{ding2022non}%
  \BibitemOpen
  \bibfield  {author} {\bibinfo {author} {\bibfnamefont {K.}~\bibnamefont {Ding}}, \bibinfo {author} {\bibfnamefont {C.}~\bibnamefont {Fang}},\ and\ \bibinfo {author} {\bibfnamefont {G.}~\bibnamefont {Ma}},\ }\bibfield  {title} {\bibinfo {title} {Non-{H}ermitian topology and exceptional-point geometries},\ }\href@noop {} {\bibfield  {journal} {\bibinfo  {journal} {Nat. Rev. Phys.}\ }\textbf {\bibinfo {volume} {4}},\ \bibinfo {pages} {745} (\bibinfo {year} {2022})}\BibitemShut {NoStop}%
\bibitem [{\citenamefont {Bergholtz}\ \emph {et~al.}(2021)\citenamefont {Bergholtz}, \citenamefont {Budich},\ and\ \citenamefont {Kunst}}]{RevModPhys.93.015005}%
  \BibitemOpen
  \bibfield  {author} {\bibinfo {author} {\bibfnamefont {E.~J.}\ \bibnamefont {Bergholtz}}, \bibinfo {author} {\bibfnamefont {J.~C.}\ \bibnamefont {Budich}},\ and\ \bibinfo {author} {\bibfnamefont {F.~K.}\ \bibnamefont {Kunst}},\ }\bibfield  {title} {\bibinfo {title} {Exceptional topology of non-{H}ermitian systems},\ }\href {https://doi.org/10.1103/RevModPhys.93.015005} {\bibfield  {journal} {\bibinfo  {journal} {Rev. Mod. Phys.}\ }\textbf {\bibinfo {volume} {93}},\ \bibinfo {pages} {015005} (\bibinfo {year} {2021})}\BibitemShut {NoStop}%
\bibitem [{\citenamefont {Ju}\ \emph {et~al.}(2022)\citenamefont {Ju}, \citenamefont {Miranowicz}, \citenamefont {Minganti}, \citenamefont {Chan}, \citenamefont {Chen},\ and\ \citenamefont {Nori}}]{PhysRevResearch.4.023070}%
  \BibitemOpen
  \bibfield  {author} {\bibinfo {author} {\bibfnamefont {C.-Y.}\ \bibnamefont {Ju}}, \bibinfo {author} {\bibfnamefont {A.}~\bibnamefont {Miranowicz}}, \bibinfo {author} {\bibfnamefont {F.}~\bibnamefont {Minganti}}, \bibinfo {author} {\bibfnamefont {C.-T.}\ \bibnamefont {Chan}}, \bibinfo {author} {\bibfnamefont {G.-Y.}\ \bibnamefont {Chen}},\ and\ \bibinfo {author} {\bibfnamefont {F.}~\bibnamefont {Nori}},\ }\bibfield  {title} {\bibinfo {title} {Einstein's quantum elevator: {H}ermitization of non-{H}ermitian {H}amiltonians via a generalized vielbein formalism},\ }\href {https://doi.org/10.1103/PhysRevResearch.4.023070} {\bibfield  {journal} {\bibinfo  {journal} {Phys. Rev. Res.}\ }\textbf {\bibinfo {volume} {4}},\ \bibinfo {pages} {023070} (\bibinfo {year} {2022})}\BibitemShut {NoStop}%
\bibitem [{\citenamefont {Ju}\ \emph {et~al.}(2024{\natexlab{a}})\citenamefont {Ju}, \citenamefont {Miranowicz}, \citenamefont {Chen}, \citenamefont {Chen},\ and\ \citenamefont {Nori}}]{ju2024emergent}%
  \BibitemOpen
  \bibfield  {author} {\bibinfo {author} {\bibfnamefont {C.-Y.}\ \bibnamefont {Ju}}, \bibinfo {author} {\bibfnamefont {A.}~\bibnamefont {Miranowicz}}, \bibinfo {author} {\bibfnamefont {Y.-N.}\ \bibnamefont {Chen}}, \bibinfo {author} {\bibfnamefont {G.-Y.}\ \bibnamefont {Chen}},\ and\ \bibinfo {author} {\bibfnamefont {F.}~\bibnamefont {Nori}},\ }\bibfield  {title} {\bibinfo {title} {Emergent parallel transport and curvature in {H}ermitian and non-{H}ermitian quantum mechanics},\ }\href {https://quantum-journal.org/papers/q-2024-03-13-1277/} {\bibfield  {journal} {\bibinfo  {journal} {Quantum}\ }\textbf {\bibinfo {volume} {8}},\ \bibinfo {pages} {1277} (\bibinfo {year} {2024}{\natexlab{a}})}\BibitemShut {NoStop}%
\bibitem [{\citenamefont {Ju}\ \emph {et~al.}(2024{\natexlab{b}})\citenamefont {Ju}, \citenamefont {Huang},\ and\ \citenamefont {Chen}}]{ju2024event}%
  \BibitemOpen
  \bibfield  {author} {\bibinfo {author} {\bibfnamefont {C.-Y.}\ \bibnamefont {Ju}}, \bibinfo {author} {\bibfnamefont {F.-H.}\ \bibnamefont {Huang}},\ and\ \bibinfo {author} {\bibfnamefont {G.-Y.}\ \bibnamefont {Chen}},\ }\bibfield  {title} {\bibinfo {title} {Event-{H}orizon-{L}ike {S}ingularities and {Q}uantum {P}hase {T}ransitions},\ }\href {https://arxiv.org/abs/2403.16503} {\bibfield  {journal} {\bibinfo  {journal} {arXiv:2403.16503}\ } (\bibinfo {year} {2024}{\natexlab{b}})}\BibitemShut {NoStop}%
\bibitem [{\citenamefont {Cainelli}\ \emph {et~al.}(2022)\citenamefont {Cainelli}, \citenamefont {Borrelli},\ and\ \citenamefont {Tanimura}}]{Tanimura03}%
  \BibitemOpen
  \bibfield  {author} {\bibinfo {author} {\bibfnamefont {M.}~\bibnamefont {Cainelli}}, \bibinfo {author} {\bibfnamefont {R.}~\bibnamefont {Borrelli}},\ and\ \bibinfo {author} {\bibfnamefont {Y.}~\bibnamefont {Tanimura}},\ }\bibfield  {title} {\bibinfo {title} {{Effect of mixed Frenkel and charge transfer states in time-gated fluorescence spectra of perylene bisimides H-aggregates: Hierarchical equations of motion approach}},\ }\href {https://doi.org/10.1063/5.0102000} {\bibfield  {journal} {\bibinfo  {journal} {J. Chem. Phys}\ }\textbf {\bibinfo {volume} {157}},\ \bibinfo {pages} {084103} (\bibinfo {year} {2022})}\BibitemShut {NoStop}%
\bibitem [{\citenamefont {Ikeda}\ and\ \citenamefont {Tanimura}(2017)}]{Tanimura04}%
  \BibitemOpen
  \bibfield  {author} {\bibinfo {author} {\bibfnamefont {T.}~\bibnamefont {Ikeda}}\ and\ \bibinfo {author} {\bibfnamefont {Y.}~\bibnamefont {Tanimura}},\ }\bibfield  {title} {\bibinfo {title} {{Probing photoisomerization processes by means of multi-dimensional electronic spectroscopy: The multi-state quantum hierarchical Fokker-Planck equation approach}},\ }\href {https://doi.org/10.1063/1.4989537} {\bibfield  {journal} {\bibinfo  {journal} {J. Chem. Phys.}\ }\textbf {\bibinfo {volume} {147}},\ \bibinfo {pages} {014102} (\bibinfo {year} {2017})}\BibitemShut {NoStop}%
\end{thebibliography}%


\begin{thebibliography}{3}%
\makeatletter
\providecommand \@ifxundefined [1]{%
 \@ifx{#1\undefined}
}%
\providecommand \@ifnum [1]{%
 \ifnum #1\expandafter \@firstoftwo
 \else \expandafter \@secondoftwo
 \fi
}%
\providecommand \@ifx [1]{%
 \ifx #1\expandafter \@firstoftwo
 \else \expandafter \@secondoftwo
 \fi
}%
\providecommand \natexlab [1]{#1}%
\providecommand \enquote  [1]{``#1''}%
\providecommand \bibnamefont  [1]{#1}%
\providecommand \bibfnamefont [1]{#1}%
\providecommand \citenamefont [1]{#1}%
\providecommand \href@noop [0]{\@secondoftwo}%
\providecommand \href [0]{\begingroup \@sanitize@url \@href}%
\providecommand \@href[1]{\@@startlink{#1}\@@href}%
\providecommand \@@href[1]{\endgroup#1\@@endlink}%
\providecommand \@sanitize@url [0]{\catcode `\\12\catcode `\$12\catcode `\&12\catcode `\#12\catcode `\^12\catcode `\_12\catcode `\%12\relax}%
\providecommand \@@startlink[1]{}%
\providecommand \@@endlink[0]{}%
\providecommand \url  [0]{\begingroup\@sanitize@url \@url }%
\providecommand \@url [1]{\endgroup\@href {#1}{\urlprefix }}%
\providecommand \urlprefix  [0]{URL }%
\providecommand \Eprint [0]{\href }%
\providecommand \doibase [0]{https://doi.org/}%
\providecommand \selectlanguage [0]{\@gobble}%
\providecommand \bibinfo  [0]{\@secondoftwo}%
\providecommand \bibfield  [0]{\@secondoftwo}%
\providecommand \translation [1]{[#1]}%
\providecommand \BibitemOpen [0]{}%
\providecommand \bibitemStop [0]{}%
\providecommand \bibitemNoStop [0]{.\EOS\space}%
\providecommand \EOS [0]{\spacefactor3000\relax}%
\providecommand \BibitemShut  [1]{\csname bibitem#1\endcsname}%
\let\auto@bib@innerbib\@empty
\bibitem [{\citenamefont {Han}\ \emph {et~al.}(2018)\citenamefont {Han}, \citenamefont {Zhang}, \citenamefont {Zheng},\ and\ \citenamefont {Yan}}]{exactruncation2018}%
  \BibitemOpen
  \bibfield  {author} {\bibinfo {author} {\bibfnamefont {L.}~\bibnamefont {Han}}, \bibinfo {author} {\bibfnamefont {H.-D.}\ \bibnamefont {Zhang}}, \bibinfo {author} {\bibfnamefont {X.}~\bibnamefont {Zheng}},\ and\ \bibinfo {author} {\bibfnamefont {Y.}~\bibnamefont {Yan}},\ }\bibfield  {title} {\bibinfo {title} {{On the exact truncation tier of fermionic hierarchical equations of motion}},\ }\href {https://doi.org/10.1063/1.5034776} {\bibfield  {journal} {\bibinfo  {journal} {J. Chem. Phys.}\ }\textbf {\bibinfo {volume} {148}},\ \bibinfo {pages} {234108} (\bibinfo {year} {2018})}\BibitemShut {NoStop}%
\bibitem [{\citenamefont {Breuer}\ and\ \citenamefont {Petruccione}(2002)}]{breuer2002theory}%
  \BibitemOpen
  \bibfield  {author} {\bibinfo {author} {\bibfnamefont {H.-P.}\ \bibnamefont {Breuer}}\ and\ \bibinfo {author} {\bibfnamefont {F.}~\bibnamefont {Petruccione}},\ }\href@noop {} {\emph {\bibinfo {title} {The theory of open quantum systems}}}\ (\bibinfo  {publisher} {Oxford University Press},\ \bibinfo {year} {2002})\BibitemShut {NoStop}%
\bibitem [{\citenamefont {Breuer}\ \emph {et~al.}(2016)\citenamefont {Breuer}, \citenamefont {Laine}, \citenamefont {Piilo},\ and\ \citenamefont {Vacchini}}]{RevModPhys.88.021002}%
  \BibitemOpen
  \bibfield  {author} {\bibinfo {author} {\bibfnamefont {H.-P.}\ \bibnamefont {Breuer}}, \bibinfo {author} {\bibfnamefont {E.-M.}\ \bibnamefont {Laine}}, \bibinfo {author} {\bibfnamefont {J.}~\bibnamefont {Piilo}},\ and\ \bibinfo {author} {\bibfnamefont {B.}~\bibnamefont {Vacchini}},\ }\bibfield  {title} {\bibinfo {title} {{C}olloquium: {N}on-{M}arkovian dynamics in open quantum systems},\ }\href {https://doi.org/10.1103/RevModPhys.88.021002} {\bibfield  {journal} {\bibinfo  {journal} {Rev. Mod. Phys.}\ }\textbf {\bibinfo {volume} {88}},\ \bibinfo {pages} {021002} (\bibinfo {year} {2016})}\BibitemShut {NoStop}%
\end{thebibliography}%

\end{document}


\title{Supplementary Information to ``Non-Markovian Quantum Exceptional Points"}
\author{Jhen-Dong Lin}
\email{jhendonglin@gmail.com}
\affiliation{Department of Physics, National Cheng Kung University, 701 Tainan, Taiwan}
\affiliation{Center for Quantum Frontiers of Research \& Technology, NCKU, 70101 Tainan, Taiwan}

\author{Po-Chen Kuo}
\affiliation{Department of Physics, National Cheng Kung University, 701 Tainan, Taiwan}
\affiliation{Center for Quantum Frontiers of Research \& Technology, NCKU, 70101 Tainan, Taiwan}

\author{Neill Lambert}
\affiliation{Theoretical Quantum Physics Laboratory, RIKEN Cluster for Pioneering Research, Wako-shi, Saitama 351-0198, Japan}
\affiliation{RIKEN Center for Quantum Computing (RQC), Wakoshi, Saitama 351-0198, Japan}
\author{Adam Miranowicz}
\affiliation{Institute of Spintronics and Quantum Information, Faculty of Physics, Adam Mickiewicz University, 61-614 Poznań, Poland}

\author{Franco Nori}
\email{fnori@riken.jp}
\affiliation{Theoretical Quantum Physics Laboratory, RIKEN Cluster for Pioneering Research, Wako-shi, Saitama 351-0198, Japan}
\affiliation{RIKEN Center for Quantum Computing (RQC), Wakoshi, Saitama 351-0198, Japan}
\affiliation{Quantum Research Institute, The University of Michigan, Ann Arbor, 48109-1040 Michigan, USA}
\author{Yueh-Nan Chen}
\email{yuehnan@mail.ncku.edu.tw}
\affiliation{Department of Physics, National Cheng Kung University, 701 Tainan, Taiwan}
\affiliation{Center for Quantum Frontiers of Research \& Technology, NCKU, 70101 Tainan, Taiwan}
\affiliation{Physics Division, National Center for Theoretical Sciences, Taipei 106319, Taiwan}



\setcounter{equation}{0}
\setcounter{figure}{0}
\renewcommand{\thefigure}{S\arabic{figure}}
\setcounter{section}{0}
\setcounter{table}{0}
\renewcommand{\thetable}{S\arabic{table}}
\setcounter{page}{1}
\maketitle

\section*{Supplementary Note 1. The extended Liouvillian superoperators for the spin-boson model}
\subsection{The pseudomode equation of motion approach}
Here, we present the extended Liouvillian superoperator of the PMEOM for the spin-boson model. Let us begin with the interaction Hamiltonian within the interaction picture, which is written as
\begin{equation}
    \begin{aligned}
        H^I_{\s\E} = \sum_k g_k \left[e^{i(\omega_0 - \omega_k)}\sigma_+ b_k + e^{-i(\omega_0 - \omega_k)}\sigma_-b_k^\dag\right].
    \end{aligned}
\end{equation}
As mentioned in the main text, we consider a parametrized spectral density, which is written as 
\begin{equation}
    J_q(\omega)=\frac{\Gamma \Lambda^2}{2[(\omega-\omega_0)^2 +\Lambda^2]}-\frac{\Gamma (q\Lambda)^2}{2[(\omega-\omega_0)^2 +(q\Lambda)^2]}.
\end{equation} 
One can then construct the following PMEOM:
\begin{equation}
    \begin{aligned}
        &\frac{d}{dt}\rho_{\s+\PM}(t)= \mathcal{L}_{\s+\PM}[\rho_{\s+\PM}(t)]
        =-i[H_{\s+\PM},\rho_{\s+\PM}(t)]+\sum_{i=1,2}\gamma_i \mathcal{L}_{a_i}[\rho_{\s+\PM}(t)],\\
        &\text{~with~}H_{\s+\PM}=\sum_{i=1,2}\alpha_i (\sigma_- a_i^\dag + \sigma_+ a_i).
    \end{aligned}
\end{equation}
Here, the qubit-pseudomode coupling strengths and the pseudomode damping rates are $\{\alpha_1 = \sqrt{\Lambda \Gamma/2}, \alpha_2 = i\sqrt{q\Lambda \Gamma/2}, \gamma_1 = 2\Lambda, \gamma_2= 2q\Lambda\}$, respectively. 

To perform spectral analysis on the extended Liouvillian superoperator, we consider the vectorization representation of the PMEOM, which can be expressed by
\begin{equation}
    \begin{aligned}
    \frac{d}{dt}\textbf{vec}[{\rho}_{\s+\PM}(t)]= \bar{\bar{\mathcal{L}}}_{\s+\PM}\textbf{vec}[{\rho}_{\s+\PM}(t)],
    \end{aligned}
\end{equation}
where $\textbf{vec}(\bullet)$ and $\bar{\bar{\mathcal{L}}}_{\s+\PM}$ denote a vectorization operation and a matrix representation of the extended superoperator, respectively. Specifically, consider an operator $A$ with a matrix representation in terms of a basis $\{\ket{i}\}$, namely, $A=\sum_{i,j}A_{i,j}\ket{i}\bra{j}$, the vectorization operation is described by $\textbf{vec}(A) = \sum_{i,j}A_{i.j}\ket{i}\otimes \ket{j}$. In this case, the matrix representation of the extended superoperator can be written as 
\begin{equation}
    \begin{aligned}
        \bar{\bar{\mathcal{L}}}_{\s+\PM} = -i( H_{\s+\PM}\otimes \mathbb{1}-\mathbb{1}\otimes H_{\s+\PM}^{T})+\sum_{i=1,2}\frac{\gamma_i}{2}(2 a_i\otimes a_i^*-a_i^\dag a_i \otimes \mathbb{1}-\mathbb{1}\otimes a_i^T a_i^* ),
    \end{aligned}
\end{equation}
where the superscripts $T$ and $*$ denote the transpose and complex conjugate operations, respectively.
Because we consider a zero-temperature environment with RWA, one can confine the analysis within the single-excitation subspace.  For the case without the band gap ($q=0$), the corresponding single-excitation subspace is spanned by $\{\ket{g,0}, \ket{e,0}, \ket{g,1}\}$, and the extended Liouvillian superoperator can be represented by a 9$\times$9 matrix:
\begin{equation}
   \bar{\bar{\mathcal{L}}}_{\s+\PM}= \begin{bmatrix}
 0 & 0 & 0 & 0 & 0 & 0 & 0 & 0 & 2 \Lambda  \\
 0 & 0 & \frac{i \sqrt{\Gamma  \Lambda }}{\sqrt{2}} & 0 & 0 & 0 & 0 & 0 & 0 \\
 0 & \frac{i \sqrt{\Gamma  \Lambda }}{\sqrt{2}} & -\Lambda  & 0 & 0 & 0 & 0 & 0 & 0 \\
 0 & 0 & 0 & 0 & 0 & 0 & -\frac{i \sqrt{\Gamma  \Lambda }}{\sqrt{2}} & 0 & 0 \\
 0 & 0 & 0 & 0 & 0 & \frac{i \sqrt{\Gamma  \Lambda }}{\sqrt{2}} & 0 & -\frac{i \sqrt{\Gamma  \Lambda }}{\sqrt{2}} & 0 \\
 0 & 0 & 0 & 0 & \frac{i \sqrt{\Gamma  \Lambda }}{\sqrt{2}} & -\Lambda  & 0 & 0 & -\frac{i \sqrt{\Gamma  \Lambda }}{\sqrt{2}} \\
 0 & 0 & 0 & -\frac{i \sqrt{\Gamma  \Lambda }}{\sqrt{2}} & 0 & 0 & -\Lambda  & 0 & 0 \\
 0 & 0 & 0 & 0 & -\frac{i \sqrt{\Gamma  \Lambda }}{\sqrt{2}} & 0 & 0 & -\Lambda  & \frac{i \sqrt{\Gamma  \Lambda }}{\sqrt{2}} \\
 0 & 0 & 0 & 0 & 0 & -\frac{i \sqrt{\Gamma  \Lambda }}{\sqrt{2}} & 0 & \frac{i \sqrt{\Gamma  \Lambda }}{\sqrt{2}} & -2 \Lambda  
\end{bmatrix}.
\end{equation}
At the EP condition $\Gamma = \Lambda/2$, one can perform a Jordan decomposition and obtain
\begin{equation}
    \begin{aligned}
        &\bar{\bar{\mathcal{L}}}_{\s+\PM}(\Gamma=\Lambda/2)= S D S^{-1}, \\ 
        &\text{with~} D=
        \begin{bmatrix}
         0 & 0 & 0 & 0 & 0 & 0 & 0 & 0 & 0 \\
         0 & -\Lambda  & 0 & 0 & 0 & 0 & 0 & 0 & 0 \\
         0 & 0 & -\Lambda  & 1 & 0 & 0 & 0 & 0 & 0 \\
         0 & 0 & 0 & -\Lambda  & 1 & 0 & 0 & 0 & 0 \\
         0 & 0 & 0 & 0 & -\Lambda  & 0 & 0 & 0 & 0 \\
         0 & 0 & 0 & 0 & 0 & -\Lambda/2 & 1 & 0 & 0 \\
         0 & 0 & 0 & 0 & 0 & 0 & -\Lambda/2 & 0 & 0 \\
         0 & 0 & 0 & 0 & 0 & 0 & 0 & -\Lambda/2 & 1 \\
         0 & 0 & 0 & 0 & 0 & 0 & 0 & 0 & -\Lambda/2
        \end{bmatrix} \\
        &\text{~and~}S=
        \begin{bmatrix}
         1 & 0 & -\Lambda ^2 & \Lambda  & -1 & 0 & 0 & 0 & 0 \\
         0 & 0 & 0 & 0 & 0 & 0 & 0 & -i & -2i \Lambda^{-1} \\
         0 & 0 & 0 & 0 & 0 & 0 & 0 & 1 & 0 \\
         0 & 0 & 0 & 0 & 0 & i & 2i \Lambda^{-1} & 0 & 0 \\
         0 & 0 & \Lambda ^2/2 & 0 & 0 & 0 & 0 & 0 & 0 \\
         0 & 1/2 & i\Lambda ^2/2 & -i\Lambda/2  & 0 & 0 & 0 & 0 & 0 \\
         0 & 0 & 0 & 0 & 0 & 1 & 0 & 0 & 0 \\
         0 & 1/2 & -i\Lambda^2/2 & i \Lambda/2  & 0 & 0 & 0 & 0 & 0 \\
         0 & 0 & \Lambda ^2/2 & -\Lambda  & 1 & 0 & 0 & 0 & 0
        \end{bmatrix}.\label{jordan_pseudo}
    \end{aligned}
\end{equation}
Therefore, by observing the matrix $D$, one can conclude that an EP3 emerges with a converged eigenvalue $-\Lambda$, and two EP2s emerge with a converged eigenvalue $-\Lambda/2$. 
For the scenario with $q>0$, where the corresponding single-excitation subspace is spanned by $\{\ket{g,0,0}, \ket{e,0,0},\ket{g,1,0}, \ket{g,0,1}\}$. Following a similar procedure, one can express the extended Liouvillian superoperator as a $16\times 16$ matrix. By performing the Jordan decomposition, one can conclude that at the EP condition $\Gamma=(1-q)\Lambda/2$, an EP3 emerges with a converged eigenvalue $-(1+q)\Lambda$ and four EP2s with a converged eigenvalue $-(1+q)\Lambda/2$.

\subsection{The hierarchical equations of motion approach}
We now present the extended Liouvillian superoperator within the HEOM formalism. We focus on presenting the results for the spin-boson model without the bandgap ($q=0$). When $q \neq 0$, a $60\times 60$ matrix is required for the extended Liouvillain superoperator,  making it impractical to present fully. Nevertheless, the procedural logic remains consistent across different $q$ values.

To facilitate spectral analysis, we begin by considering the vectorized representation of the HEOM, which can be expressed as
\begin{equation}
    \begin{aligned}
    \frac{d}{dt}\textbf{vec}[{\rho}_{\s+\ADO}(t)]= \bar{\bar{\mathcal{L}}}^{\text{RWA}}_{\s+\ADO}\textbf{vec}[{\rho}_{\s+\ADO}(t)],
    \end{aligned}
\end{equation}
where $\bar{\bar{\mathcal{L}}}^{\text{RWA}}_{\s+\ADO}$ is given by
\begin{equation}
\begin{aligned}
\bar{\bar{\mathcal{L}}}_{\text{S}+\text{ADO}}^{\text{RWA}}\textbf{vec}[\rho^{(m)}_{\bfk}(t)]
=\bar{\bar{\mathcal{L}}}_{0}\textbf{vec}[\rho^{(m)}_{\bfk}(t)]
-\sum_{r=1}^{m}\chi_{k_{r}}\textbf{vec}[\rho^{(m)}_{\bfk}(t)]
-i\sum_{k'} \bar{\bar{\mathcal{A}}}_{k'}
\textbf{vec}[\rho^{(m+1)}_{\bfk^+}(t)]
-i\sum_{r=1}^{m}\bar{\bar{\mathcal{B}}}_{k_{r}}
\textbf{vec}[\rho^{(m-1)}_{\bfk_{r}^{-}}(t)].
\label{eq:HEOML_RWA1}
\end{aligned}
\end{equation}%
The required superoperators acting on the vectorized ADOs are
\begin{equation}
    \begin{aligned}
        \bar{\bar{\mathcal{L}}}_{0} = -i~(H_{\s}\otimes \mathbb{1}-\mathbb{1}\otimes H_{\s}^{T}),
    \end{aligned}
\end{equation}
\begin{equation}
\begin{aligned}
\bar{\bar{\mathcal{A}}}_{k}[\cdot] = \sigma_{\bar{\nu}}\otimes \mathbb{1}-\mathbb{1}\otimes\sigma_{\nu}
\label{eq:A1}
\end{aligned}
\end{equation}%
and
\begin{equation}
\begin{aligned}
\bar{\bar{\mathcal{B}}}_{k}[\cdot] = \xi^{\nu}_{l}~\sigma_{\nu}\otimes \mathbb{1}-\xi^{\bar{\nu}\ast}_{l}~\mathbb{1}\otimes\sigma_{\bar{\nu}},
\label{eq:B1}
\end{aligned}
\end{equation}%
where $H_\s=\omega_0 |e\rangle \langle e|$. Due to the absence of internal tunneling between the qubit energy levels, the system's behavior can be accurately captured by considering up to the second hierarchical tier ($m=2$)~\cite{exactruncation2018}. The validity of this truncation can be further verified using Eq. (15) of the main text. The corresponding extended Liouvillian superoperator can be represented as a $24\times24$ matrix. To facilitate analytical calculations, we decompose $\bar{\bar{\mathcal{L}}}_{\text{S}+\text{ADO}}^{\text{RWA}}$ into three components
\begin{equation}
\begin{aligned}
\bar{\bar{\mathcal{L}}}_{\text{S}+\text{ADO}}^{\text{RWA}}
=\bar{\bar{\mathcal{L}}}_{\text{S}+\text{ADO}}^{\text{RWA},\text{p}}
\oplus\bar{\bar{\mathcal{L}}}_{\text{S}+\text{ADO}}^{\text{RWA},\text{c}}
\oplus\bar{\bar{\mathcal{L}}}_{\text{S}+\text{ADO}}^{\text{RWA},\text{c}\ast},
\label{eq:HEOML_RWA2}
\end{aligned}
\end{equation}%
where the superscripts p, c, and $\text{c}\ast$ respectively indicate that these components determine the system's population (p), coherence (c), and its complex conjugate ($\text{c}\ast$). The full expressions for these components are given by
\begin{equation}
    \begin{aligned}
    &\bar{\bar{\mathcal{L}}}_{\text{S}+\text{ADO}}^{\text{RWA},\text{p}}=
\begin{bmatrix}
 0 & 0 & -i & i & 0 & 0 \\
 0 & 0 & i & -i & 0 & 0 \\
 0 & i \Gamma  \Lambda/2 & -\Lambda  & 0 & -i & i \\
 0 & -i \Gamma  \Lambda/2  & 0 & -\Lambda  & i & -i \\
 0 & 0 & -i \Gamma  \Lambda/2  & i \Gamma  \Lambda/2 & -2 \Lambda  & 0 \\
 0 & 0 & 0 & 0 & 0 & -2 \Lambda  \\
\end{bmatrix}, \label{Lp}
    \end{aligned}
\end{equation}%

\begin{equation}
    \begin{aligned}
    &\bar{\bar{\mathcal{L}}}_{\text{S}+\text{ADO}}^{\text{RWA},\text{c}}=
\begin{bmatrix}
 0 & -i & i & 0 & 0 \\
 - i \Gamma  \Lambda/2  & -\Lambda  & 0 & -i & i \\
 0 & 0 & -\Lambda  & i & -i \\
 0 & 0 & i \Gamma  \Lambda/2 & -2 \Lambda  & 0 \\
 0 & 0 & -i \Gamma  \Lambda/2  & 0 & -2 \Lambda  \\
\end{bmatrix}, \label{Lc}
    \end{aligned}
\end{equation}%
and 
\begin{equation}
    \begin{aligned}
  &\bar{\bar{\mathcal{L}}}_{\text{S}+\text{ADO}}^{\text{RWA},\text{c}\ast}=
\begin{bmatrix}
 0 & i & -i & 0 & 0 \\
 i \Gamma  \Lambda/2 & -\Lambda  & 0 & -i & i \\
 0 & 0 & -\Lambda  & i & -i \\
 0 & 0 & i \Gamma  \Lambda/2 & -2 \Lambda  & 0 \\
 0 & 0 & -i \Gamma  \Lambda/2  & 0 & -2 \Lambda  \\
\end{bmatrix}.\label{Lcc}
    \end{aligned}
\end{equation}%
Under the EP condition ($\Gamma = \Lambda/2$), we perform Jordan decomposition and  obtain
\begin{equation}
    \begin{aligned}
    \bar{\bar{\mathcal{L}}}_{\text{S}+\text{ADO}}^{\text{RWA},\text{p}}&= S^{(\text{p})} D^{(\text{p})} S^{(\text{p})-1}, \\ 
        \text{with~} D^{(p)}&=
        \begin{bmatrix}
         0 & 0 & 0 & 0 & 0 & 0 \\
         0 & -2 \Lambda  & 0 & 0 & 0 & 0 \\
         0 & 0 & -\Lambda  & 0 & 0 & 0 \\
         0 & 0 & 0 & -\Lambda  & 1 & 0 \\
         0 & 0 & 0 & 0 & -\Lambda  & 1 \\
         0 & 0 & 0 & 0 & 0 & -\Lambda  \\
        \end{bmatrix} \\ 
        \text{~and~}S^{(\text{p})}&=
        \begin{bmatrix}
         1 & 0 & 0 & -2 & 0 & 0 \\
         0 & 0 & 0 & 2 & 0 & 0 \\
         0 & 0 & 1/2 & i \Lambda  & -i & 0 \\
         0 & 0 & 1/2 & -i \Lambda  & i & 0 \\
         0 & 1 & 0 & \Lambda ^2/2 & -\Lambda  & 1 \\
         0 & 1 & 0 & 0 & 0 & 0 \\
        \end{bmatrix},\\ \\
\bar{\bar{\mathcal{L}}}_{\text{S}+\text{ADO}}^{\text{RWA},\text{c}}&= S^{(\text{c})} D^{(\text{c})} S^{(\text{c})-1}, \\ 
        \text{with~} D^{(\text{c})}&=
        \begin{bmatrix}
         -2 \Lambda  & 0 & 0 & 0 & 0 \\
         0 & (-3-i)\Lambda/2  & 0 & 0 & 0 \\
         0 & 0 & (-3+i)\Lambda/2  & 0 & 0 \\
         0 & 0 & 0 & -\Lambda/2 & 1 \\
         0 & 0 & 0 & 0 & -\Lambda/2 \\
        \end{bmatrix} \\
        \text{~and~}S^{(\text{c})}&=
        \begin{bmatrix}
         0 & (48-64i)/25\Lambda^2 & (48+64i)/25\Lambda^2 & 2i/\Lambda  & 4i/\Lambda ^2 \\
         0 & (-22-54i)/25\Lambda & (22-54i)/25\Lambda & 1 & 0 \\
         0 & (2+2 i)/{\Lambda } & -({2-2 i}){\Lambda } & 0 & 0 \\
         1 & -1 & -1 & 0 & 0 \\
         1 & 1 & 1 & 0 & 0 \\
        \end{bmatrix},\\ \\
        \bar{\bar{\mathcal{L}}}_{\text{S}+\text{ADO}}^{\text{RWA},\text{c}\ast}&= S^{(\text{c}\ast)} D^{(\text{c}\ast)} S^{(\text{c}\ast)-1}, \\ 
        \text{with~} D^{(\text{c}\ast)}&=
        \begin{bmatrix}
         -2 \Lambda  & 0 & 0 & 0 & 0 \\
         0 & (-3-i)\Lambda/2  & 0 & 0 & 0 \\
         0 & 0 & (-3+i)\Lambda/2  & 0 & 0 \\
         0 & 0 & 0 & -\Lambda/2 & 1 \\
         0 & 0 & 0 & 0 & -\Lambda/2 \\
        \end{bmatrix} \\
        \text{~and~}S^{(\text{c}\ast)}&=
        \begin{bmatrix}
         0 & (48-64i)/25\Lambda ^2 & (-48-64i)/25\Lambda ^2 & -2i/\Lambda & -4i/{\Lambda ^2} \\
         0 & (-22-54i)/25\Lambda & (22-54i)/25\Lambda & 1 & 0 \\
         0 & (2+2 i)/{\Lambda } & (-2+2 i)/{\Lambda } & 0 & 0 \\
         1 & -1 & -1 & 0 & 0 \\
         1 & 1 & 1 & 0 & 0 \\
        \end{bmatrix}.
    \end{aligned}
\end{equation}
By observing the matrices $D^{(\text{p}),(\text{c}),(\text{c}\ast)}$, we conclude the emergence of one EP3 with a converged eigenvalue $-\Lambda$, and two EP2s with a converged eigenvalue $-\Lambda/2$. This aligns with the results obtained using the PMEOM approach. Moreover, we conclude that the EP3 can be observed through the system's population dynamics, whereas the two EP2s can be observed through the system's coherence.
We note that within the HEOM formalism, there are additional eigenmatrices introduced that do not influence the system dynamics due to the vanishing of the corresponding coefficients $c_i$ and $c_m$ as described in Eq. (15) in the main text.

\section*{Supplementary Note 2. Markovian-to-non-Markovian transition for the spin-boson model}
In this section, we present the transition point from Markovian to non-Markovian dynamics in the spin-boson model, where the analytical solution for the qubit reduced state is expressed as~\cite{breuer2002theory}
\begin{equation}
\begin{aligned}
    &\bra{e}\rho_\s(t)\ket{e} = \bra{e}\rho_\s(0)\ket{e}|G(t)|^2~\text{and~}\bra{g}\rho_\s(t)\ket{e} =\bra{g}\rho_\s(0)\ket{e} G(t).
    \label{eq: analytical solution}
\end{aligned}
\end{equation}
Here, $G(t)$ is termed as the decoherence function, which is given by
\begin{equation}
    \begin{aligned}
        &
    G(t)=\frac{e^{-\frac{1}{2}\Lambda  \delta_+ t }}{\Gamma\delta_-+2 \Lambda  q}\Biggl\{ 
 2 \Lambda  q \cosh \left(\frac{\Lambda  \delta_+ t}{2} \right) 
 + \delta_-\Gamma\cosh \left(\frac{1}{2}t \sqrt{\Lambda  \delta_-(\delta_-\Lambda - 2\Gamma)}  \right)
 +2 \Lambda  q \sinh \left(\frac{\Lambda  \delta_+ t}{2} \right)\\
 &~~~~~~~~~~~~~~~~~~~~~~~~~~~~+\frac{\delta_+i\sqrt{\Lambda \delta_-} \sinh \left(\frac{1}{2}t \sqrt{\Lambda  \delta_-(\delta_-\Lambda - 2\Gamma)}\right)}{\sqrt{2 \Gamma -\Lambda  \delta_-}}
 \Biggr\}~~\text{with~~}\delta_\pm=1\pm q.
    \end{aligned} \label{eq: analytical solution}
\end{equation}
It is well-established that for both the Breuer-Lain-Pillo and the Rivas-Huelga-Plenio measures, the transition from Markovian to non-Markovian dynamics occurs when $|G(t)|$ shifts from pure decay to oscillatory~\cite{RevModPhys.88.021002}. Consequently, the transition point can be identified from Eq.~\eqref{eq: analytical solution} as $\delta_-\Lambda=2\Gamma$, or equivalently $\Gamma=(1-q)\Lambda/2$, precisely aligning with the EP condition mentioned in the main text.
\bibliography{ref_supp}